\documentstyle[aps,prb,eqsecnum,epsfig]{revtex}

\begin{document}

\def\bea{\begin{eqnarray}}
\def\eea{\end{eqnarray}}

\title{{\bf Low-energy behavior of the spin-tube and spin-orbital models}}

\author{E. Orignac$^{\dagger}$ , R. Citro$^{\star}$ and N. Andrei}

\address{{\it Serin Laboratory, Rutgers University, P.O. Box 849,
Piscataway, NJ 08855-0849, USA} \\
{\it $^{\star}$Dipartimento di Scienze Fisiche ``E.R. Caianiello'' and Unit\`a
INFM di Salerno}\\
{\it Universit\`a di Salerno, I-84081 Baronissi (Sa), Italy}}

\date{\today}

\maketitle

\begin{abstract}
The low-energy effective Hamiltonian
 of three coupled spin chains with periodic boundary conditions (spin
 tube)   
is expressed, in the limit of strong interchain coupling, in terms of 
 XXZ chains coupled by biquadratic exchange
interaction. A similar effective  model was also proposed to describe the
 coupling of spins to orbital degrees of freedom in materials such as
 $\mathrm{NaV_2O_5}$.  We investigate the effective model
  by means of bosonization and
renormalization group techniques, and
 find  that the generic phase diagram comprises a gapless region and
 gapped regions consisting of a  spin liquid phase and various
antiferromagnetic phases. We discuss the properties of the spin liquid 
phase, in particular the nature of the ground state and of the
 elementary excitations above it. 
We then study the effect of a magnetic field, and conclude
that a strong enough magnetic field can suppress the dimerized
phase leading to a two component Luttinger liquid. 
The critical
exponents at the transition gapful-gapless are calculated and shown
to be  \emph{non-universal} in the spin tube case or
 the generic spin orbital problem. 
\end{abstract}

\vspace*{1.5truecm}

\noindent PACS numbers:75.10.Jm, 75.30.Kz, 75.40.Gb

\noindent {\it Keywords}: spin-tube, bosonization and renormalization group,
spin-liquid


\section{Introduction}
Coupled spin 1/2 chains have attracted much attention lately
due to  the large
number of their experimental realizations, as well as to
the variety of
theoretical techniques, both analytical and numerical, available to
study the relevant models.
Spin 1/2 chains, owing to the Jordan-Wigner
transformation\cite{affleck_houches}, show properties
remarkably similar to those
of interacting one dimensional fermions with their low energy properties
described by an effective Luttinger liquid theory\cite{luther_chaine_xxz,haldane_xxzchain}. 
Recently it has
been realized that these properties are drastically modified when the
 spin 1/2 chains are coupled together forming ladder
systems\cite{dagotto_2ch_review}. In this case, in a way very similar to 
Haldane's spin-S problem \cite{haldane_gap}, a gap is found to open for an
even number of chains while the system remains gapless if the
number of chains is odd. This phenomenon has been thoroughly investigated
both analytically \cite{strong_spinchains}and numerically
\cite{white_2ch}, and corresponding experimental systems were identified.
Typical examples of
two-chain spin 1/2 ladders exhibiting a gap are
$SrCu_2O_3$\cite{azuma_srcuo} and
$Cu_2(C_5H_{12}N_2)_2Cl_4$\cite{chaboussant_cuhpcl}. On the other hand, an
example of a gapless
three-chain ladder is $Sr_2Cu_3O_5$\cite{azuma_srcuo}. From the
theoretical point of view, the difference between odd and even number
of legs\cite{rojo} or odd and even spin\cite{affleck_lieb}
has been understood qualitatively through a
generalization of the Lieb-Schultz-Mattis Theorem\cite{lsm_theorem,%
affleck_lsm,rojo,oshikawa}. The theorem states that if the ground state 
is unique the system will necessarily be gapless. This is indeed the case 
for  an odd number of coupled spin 1/2 chains.

More recent theoretical work has emphasized the role of  boundary conditions in
the transverse direction in the formation of
spin gaps. The results quoted above - the opening of a gap 
for an even number of coupled chains, gaplessness
 if the number is odd - are valid for 
open boundary conditions (OBC).
It turns out that when periodic boundary conditions (PBC)
are imposed on the transverse direction a gap  opens for both cases of
even and odd number of chains
although the underlying reasons and the natures of the gaps are
different. In the case of an even number of coupled chains, the reason
for the gap is the formation of spin singlets along the
transverse direction, similarly to the case of chains with open
boundary conditions. When the number of coupled chains is odd 
a two fold degenerate dimerized ground state is obtained
in the case of PBC\cite{kawan1} - in contrast to its uniqueness
in the OBC -
allowing for a gap in the spin excitations. The degeneracy of the ground
state in the case of PBC can be understood as a consequence of the
fact that PBC are frustrating for an odd number of legs. To date, no
experimental system described by coupled antiferromagnetic spin 1/2
chains with periodic boundary conditions and an odd
number of legs has been reported.

When, further,  leg-leg biquadratic interactions
are included new states emerge.
A system of  two spin-1/2 spin chains enters a spontaneously
dimerized phase with a
gapped spectrum  exhibiting non-Haldane spin-liquid
properties\cite{nersesyan_biquad}. The elementary excitations
 are neither spinons nor magnons, but pairs of
propagating triplet or singlet solitons connecting two spontaneously
dimerized ground states. Subsequently more
 non-Haldane spin-liquid models have been proposed \cite{mikeska}.
Recently, these models of spin ladders with biquadratic exchange have
been advocated\cite{mostovoy} as possible models for the formation of
a spin gap in
$\mathrm{NaV_2O_5}$ and $\mathrm{Na_2Ti_2Sb_2O}$.

\begin{figure}
\centerline{\epsfig{file=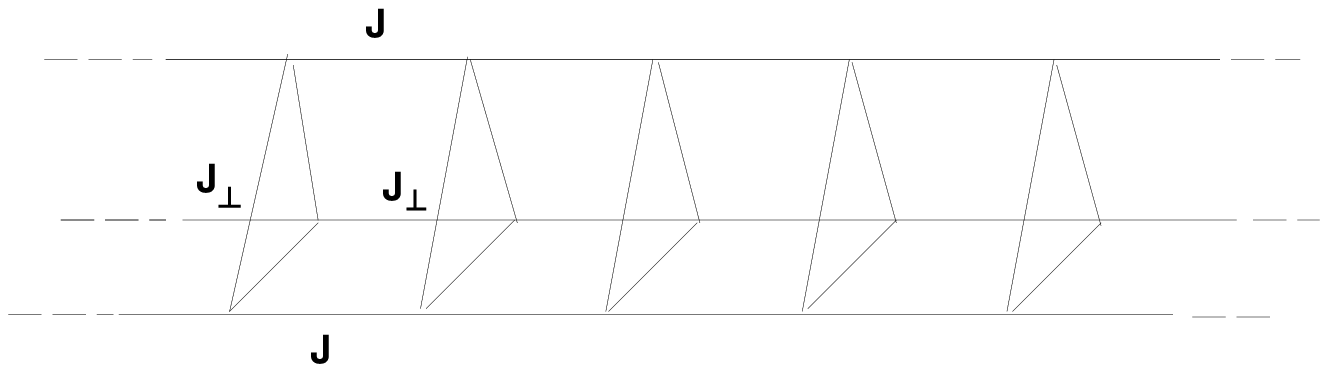,angle=0,width=12cm}}
\vspace{0.5 cm}
\caption{Cylindrical three-leg ladder (spin-tube). The choice of the
topology affects the strong-coupling limit.}
\label{fig:spin_tube}
\end{figure}

 In this paper we  study the low energy physics of the three-leg
ladder with periodic boundary conditions (see
Fig. \ref{fig:spin_tube}).
This model is
called  the spin-tube model in the following.
 By analyzing in detail
its effective  low energy Hamiltonian (LEH), which
consists of two {\it non-equivalent} coupled XXZ
chains of spins and chiral degrees of freedom, in the presence of a
biquadratic exchange\cite{schulz_berlin,kawan1}, we will show that the
dimerized ground state of this model falls in the universality class
of the non-Haldane spin liquids.
In the case of XXZ chains with a biquadratic coupling and for
$J_z/J>1$,  we find that the Ising antiferromagnetic
order can compete with the dimer order, and we will describe the
resulting phase diagram.

 The organization of the paper is as follows. In
Sec.\ref{model} we introduce the spin tube model. We sketch the
derivation of the strong interchain coupling effective Hamiltonian
 and discuss its
relation with two spin 1/2 chains coupled by a biquadratic
interaction. We then discuss the symmetries and recall the results
already known on the two chains with biquadratic interactions, in
particular the prediction of a spin gap. 

In Sec. \ref{bosonization} we discuss the bosonization treatment of
two non-equivalent XXZ chains coupled by a biquadratic interaction.
We show that this Hamiltonian contains a term $\sin 2\phi_1 \sin
2\phi_2$ that is responsible for the formation of a spin-gap and
singlet order, as well as terms $\cos 4\phi_{1,2}$ that cause
antiferromagnetic order. To analyze the competitions of these terms,
we derive renormalization group equations. Using these equations, we
estimate in which part of the phase diagram we should expect a
competition of antiferromagnetism and dimer order. We also estimate
the equation of the phase boundary between the singlet and the
Antiferromagnetic order.  

In
Sec.\ref{sec:dimerized} we
analyze in details the dimerized phase. The dimerized ground state
is two-fold degenerate and 
is formed of alternating singlets of spins and pseudospins. The
elementary excitations above the ground state carry a spin $\pm 1/2$
as well as a pseudospin $\pm 1/2$. They can be seen as formed by the
introduction of a spin and a pseudospin in the pattern of alternating
singlets. Similar excitations were obtained in
Ref.~\onlinecite{nersesyan_biquad} and it was shown that they lead to
response functions very different from those of a Haldane spin
liquid. Numerical evidence for such excitations was also obtained in
Ref. \onlinecite{cabra_unp}. 

In section \ref{sec:gap_closure}, we analyze the effect of a magnetic
field strong enough to close the gap of the dimer phase. When both
chains are magnetic, the situation is similar to the one obtained with
two coupled spin ladders \cite{cithra,furusaki_dynamical_ladder}. 
In the more physical situation, when only one of
the two chains is magnetic, the ground state properties are those of a
two component Luttinger liquid. In the case of spin-orbital models,
spin and orbital degrees of freedom decouple completely but the
presence of orbital modes should affect the specific heat. In the case
of the spin-tube, we show that in contrast to spin-orbital models, the
spin-correlation functions are affected by the presence of auxiliary
gapless modes. We also show that for non-equivalent chains (a case
that is realized in spin-orbital models) the exponent of the spin
correlations is \emph{non-universal} at the transition.
Finally, in
Sec.\ref{conclusions} we will give some concluding remarks.

\section{The spin-tube model and coupled XXZ chains}\label{model}
\subsection{The three chain ladder with periodic boundary conditions}

We wish to study the three-chain ladder
Hamiltonian,

\begin{equation}
\label{ladder}
H=J\sum_{i=1}^N \sum_{p=1}^3 \vec S_{i,p} \vec S_{i+1,p}+J_{\perp}
\sum_{i=1}^N \sum_{p=1}^3 \vec S_{i,p} \vec S_{i,p+1},
\end{equation}
\noindent where $p$ (resp. $i$) is a chain (resp. site) index, $J$ is
the coupling along the chain and $J_{\perp}$ the transverse
coupling. We impose periodic boundary conditions along the rungs by
identifying $S^\alpha_{i,4} = S^\alpha_{i,1}$.
We call the resulting model the spin-tube (see
Fig.~\ref{fig:spin_tube}) since it can be realized by placing 3 spin
1/2 chains forming an equilateral triangle.
In order to
investigate the low-energy
physics of the spin-tube, we consider the limit of strong
interchain coupling
($J_\perp \gg J$). This is the appropriate starting point that
yields a good effective description of the properties of the spin tube
in the whole range of $J_\perp/J$. (Starting from the
 opposite limit $J_\perp \ll
J$, and treating $J_\perp$  as a perturbation, one finds\cite{totsuka_3legs} 
that it gives rise to relevant terms in the
Hamiltonian. As a result, the initial $J_\perp$ grows until it is of
order of $J$ at which point the weak coupling bosonization scheme
is no more valid.)

To derive the effective low-energy Hamiltonian, let us first consider
the case $J=0$. The system is a collection of independent rungs, each
described by the following Hamiltonian
\begin{equation}
H=J_\perp
(\vec{S}_1.\vec{S}_2+\vec{S}_2.\vec{S}_3+\vec{S}_1.\vec{S}_3)=J_\perp
((\vec{S}_1+\vec{S}_2+\vec{S}_3)^2-9/4)/2.
\end{equation}

\noindent The ground state of each rung is four-fold
degenerate, composed of two doublets of spin 1/2 excitations,
corresponding to the left and right chirality (-/+), with energy $(-3/4
J_{\perp})$\cite{kawan1,schulz_houches_revue}, and
the excited states form a spin 3/2
quadruplet with energy $3/4J_{\perp}$.

Turning on $J>0$ allows the rungs to exchange spins. In the limit
$J_\perp \gg J$, the quadruplet of S=3/2 excitations can be neglected,
and only the degenerate low energy subspace of spin 1/2 states needs to be
taken into account. In this subspace the Hamiltonian transforms into
an effective Hamiltonian
with a biquadratic coupling between the spin and chirality degrees of
freedom\cite{kawan1,totsuka_3legs,schulz_moriond},

\begin{equation}
\label{heff}
H_{eff}=\frac{J}{3}\sum_i \vec S_i \vec S_{i+1} \lbrack 1+2
(\tau_i^x\tau_{i+1}^x +\tau_i^y\tau_{i+1}^y) \rbrack,
\end{equation}

\noindent where $S$ is the total spin
, and $\tau_i$ are 
operators exchanging left and right chiralities.
The original spin operators can be expressed in terms of the effective
 spin $S$ and
the chirality $\tau$ in the following way:
\begin{eqnarray}
\label{eq:spin_stot_chirality}
S_{i,p}^+=-\frac 1 3 S_i^+ + \frac 2 3 j^{2p} S_i^+ \tau_i^- + \frac 2 3 j^p
S_i^+ \tau_i^+ \nonumber \\
S_{i,p}^z=\frac 1 3 S_i^z + \frac 2 3 j^{2p} S_i^z \tau_i^- + \frac 2 3 j^p
S_i^z \tau_i^+
\end{eqnarray}
The spin tube model has already been investigated
numerically\cite{kawan1} using DMRG in the case of coupled XXX
chains. The system was shown to exhibit a spin gap $\Delta=0.28 J$,
with exponentially decaying correlation functions $(-)^n \langle S_0^z
S_n^z \rangle$ and $(-)^n \langle \tau_0^z
\tau_n^z \rangle$. This behavior was shown to be related to the
formation of a dimer order. A qualitative discussion of the origin of
this dimer order can be found in Ref. \cite{schulz_moriond}. In
Ref. \onlinecite{kawan1} the dispersion relation of excitations having
$\tau^z=0,1$ and $S=0,1$ as a function of momentum was obtained numerically for
a system of 12 sites showing only gapped excitations.   

In Ref. \onlinecite{cabra_unp}, a generalization of the effective
Hamiltonian Eq. (\ref{heff}) to coupled
anisotropic spin chains has been derived (see Eq. (\ref{eq:heff_anisotropic}).
Numerical diagonalizations were performed. A gap to $S^z=1$
excitations was obtained for$0<J_z/J<1.2$ with results in good
agreement with those of Ref. \onlinecite{kawan1} for $J_z=J$. It was
also shown that the ground state was degenerate in agreement with the
dimer order picture of Ref. \onlinecite{kawan1}. The dispersion
relation of excitations having $\tau^z=0$ and total spin $S=0$ or
$S=1$ was obtained\cite{cabra_unp}. 
It was shown to be the bottom of a two particle
continuum. The fundamental particle, the spinon, was conjectured to
have $\tau^z=\pm 1/2,S=1/2$. The spinon dispersion relation was
obtained numerically by considering a system with an odd number of
sites\cite{cabra_unp}, showing that the spinons were massive for all
momentum.  

\subsection{Generalized spin-tube model and XXZ chains coupled by a
biquadratic exchange}

The Hamiltonian (\ref{heff}) is
part of the class of the Hamiltonians consisting of
 two  non-equivalent coupled XXZ chains
in the presence of a biquadratic exchange,
\begin{equation}\label{eq:basic_hamiltonian}
H=H_0+H_{B},
\end{equation}
\noindent where
\bea
H_0 &= & \sum_i \sum_{\alpha=1,2} J_{\alpha}
(S_{\alpha,i}^xS_{\alpha,i+1}^x+S_{\alpha,i}^yS_{\alpha,i+1}^y) +
J_{\alpha}^z
S_{\alpha,i}^zS_{\alpha,i+1}^z \label{h0} \\
H_{B}&=& \lambda \sum_i (S_{1,i}^xS_{1,i+1}^x+S_{1,i}^yS_{1,i+1}^y+
\Delta_1 S_{1,i}^zS_{1,i+1}^z) (S_{2,i}^xS_{2,i+1}^x+S_{2,1}^yS_{2,i+1}^y+
\Delta_2 S_{2,i}^zS_{2,i+1}^z).\label{hbiq}
\eea
\noindent In the case of the spin-tube, $S_{1,i}$ corresponds to  spin $S_i$,
and $S_{2,i}$ is  associated with the chiral degrees of
freedom $\tau_i$. Another way of writing a class of Hamiltonian
generalizing the spin tube model  (\ref{heff}) is:

\begin{eqnarray}
H = \sum_{i=1}^{N} & &\lbrack
u+\gamma(S_{i}^{+}S_{i+1}^{-}+S_{i}^{-}S_{i+1}^{+})+J_z S_{i}^{z}
S_{i+1}^{z}\rbrack  \lbrack
v+\alpha(\tau_{i}^{+}\tau_{i+1}^{-}+\tau_{i}^{-}\tau_{i+1}^{+})+J_{z}^{\prime
} \tau_{i}^{z} \tau_{i+1}^{z}\rbrack.\label{ham}
\end{eqnarray}

The effective Hamiltonian for the spin tube
is obtained for,

\begin{eqnarray}
u=0  \; \; \gamma=J/6 \; \;J_z=J/3 \nonumber \\
v=1  \; \;  \alpha=1  \; \; J_z^\prime=0.
\end{eqnarray}

For a tube  made of  XXZ chains, one has,
instead\cite{cabra_unp}:
\begin{eqnarray}\label{eq:heff_anisotropic}
u=0 \; \; \gamma=J/6 \; \; J_z=J/3 \Delta \nonumber \\
v=1  \; \;  \alpha=1  \; \; J_z^\prime=0.
\end{eqnarray}
The parameters
$J_z/\gamma$ and $J_{z}^{\prime}/\alpha$ in Eq. (\ref{ham})
 measure the XXZ anisotropy for spin
and chirality, respectively. When both of them are equal to 1, the
Hamiltonian Eq. (\ref{ham}) is  $SU(2) \times SU(2)$ symmetric.
For $u=v$, $\alpha=\gamma$, $J_z=J_z^\prime$ the two chains are
equivalent, and can be parameterized as:

\begin{equation}\label{eq:equiv_chains_depp}
u=v=\frac{p}{2},\mbox{ }\mbox{ } \alpha=\gamma=1,\mbox{ }\mbox{ }
J_z=J_z^{'}=2q.
\end{equation}

Hamiltonians of the type Eq. (\ref{ham}) can be mapped onto
Hamiltonians of the type Eq. (\ref{eq:basic_hamiltonian}). The
correspondence is given by:
\begin{eqnarray}
J_1=2 u \alpha & &\mbox{ } \mbox{ }  J_2=2v \gamma \nonumber \\
J_1^z= v J_z   & &\mbox{ } \mbox{ }  J_2^z=u J_z^\prime \nonumber \\
\Delta_1=\frac{J_z}{2\gamma} & & \mbox{ } \mbox{ }
\Delta_2=\frac{J_z^\prime}{2\alpha}\nonumber \\
\lambda = 4 \alpha \gamma & &
\end{eqnarray}
with the identification $\vec{S}_1 \equiv \vec{S}$ , $\vec{S}_2 \equiv
\vec \tau $.
Since writing the Hamiltonian in the form  (\ref{eq:basic_hamiltonian})
 is less
restrictive than  in the 
form (\ref{ham}) (i.e.  the former 
includes the case $\Delta_\alpha \ne \frac{J_\alpha^z}{J_\alpha}$),
 we  focus on the Hamiltonian (\ref{eq:basic_hamiltonian}).
Hamiltonians of the type Eq. (\ref{eq:basic_hamiltonian}) are also 
encountered in a different context than the spin-tube model.
In particular, a Hamiltonian of the type (\ref{eq:basic_hamiltonian}) has  been
 proposed  by
Mostovoy and Khomskii as a model for the spin gap formation
in the $\mathrm{NaV_2O_5}$
\cite{mostovoy} and  $\mathrm{Na_2Ti_2Sb_2O}$ \cite{pati_orbital_dmrg}
compounds . In that case, the $S_1$ spins correspond to the real 
spin of
the system whereas the $S_2$ spins are pseudospins associated with
{\it orbital} degrees of freedom. These spin-orbital models can be
derived from a multiband Hubbard model. The derivation is reviewed for
instance in Ref. \onlinecite{khomskii}.

let us discuss first the case $J_1=J_2=J_1^z=J_2^z=J$,
$\Delta_1=\Delta_2=1$. The Hamiltonian describes two coupled
Heisenberg chains with a biquadratic coupling preserving the $SU(2)$
symmetry. Actually, the full symmetry group is larger than SU(2). One has:
\begin{equation}
[H,\vec{S}_{1,\text{tot.}}]=0 \hspace{0.5cm} [H,\vec{S}_{2,\text{tot.}}]=0
\end{equation}
and the full symmetry group is therefore
$SU(2) \times SU(2)$ rather than the SU(2) symmetry
that follows from
$[H,\vec{S}_{1,\text{tot.}}+\vec{S}_{2,\text{tot.}}]=0$.
As a result the spectrum consists of $SU(2)\times SU(2) \sim SO(4)$
multiplets\cite{nersesyan_biquad}. 
For $J=\frac \lambda 4$, the Hamiltonian (\ref{eq:basic_hamiltonian})
has been shown to have an even larger $SU(4)$ symmetry and to reduce to an
integrable $SU(4)$ spin chain \cite{deppeler}. The spectrum of the
SU(4) spin
chain has been obtained by the Bethe Ansatz \cite{uimin,sutherland},
and the correlations functions have been obtained by non-abelian
bosonization techniques\cite{aff_wz}, identifying
the low energy effective theory describing
the  spin chain as the $SU(4)_1$ WZNW model.
The integrable $SU(4)$ spin-chain has also been intensively
studied numerically using Quantum Monte Carlo (QMC) \cite{frischmuth_su4} or
Density Matrix Renormalization Group (DMRG) and Lanczos Exact
Diagonalization (ED)\cite{yamashita_su4} in the context of the
spin-orbital models. The ground state energy, and
excited state energy were obtained in good agreement\cite{yamashita_su4}
with analytical
calculation using the Bethe Ansatz\cite{sutherland}.  
The numerical calculation
of the correlation functions \cite{frischmuth_su4,yamashita_su4}
reproduces results of the  the continuum field theoretical
treatment \cite{aff_wz}.
At $J \ne \lambda/4$  perturbations are generated which lower the
$SU(4)$ symmetry to  $SU(2)\times SU(2)$ and render the Hamiltonian
non-integrable. These perturbations
have been recently studied \cite{azaria_su4} by describing the chain
away from the integrable point as a perturbed $SU(4)_1$ WZW model. 
It was found that  for $J<\lambda/4$, a gapless phase is obtained while 
for $J>\frac \lambda 4$, a gap is formed. A different field
theoretical treatment in the limit $\lambda \ll J$ also leads to the
appearance of a gapped phase. For  $J=3\lambda/4$,
the ground state wavefunction could be obtained exactly in matrix
product form\cite{mikeska},  of singlet states along the
legs of the ladder. This picture is in good agreement with the
predictions of the field theoretical treatment.  
 The dependence of the gap on the coupling for the range
 $0<\lambda/J<4$ was obtained by DMRG 
calculations\cite{pati_orbital_dmrg,yamashita_orbital_dmrg}. It was found
that 
for $\lambda/J \ll 1$,
 the gap increases proportionally to $\lambda$ in agreement with the
 weak coupling bosonization treatment, and
 vanishes for
 $\lambda/J=1/4$, which is the SU(4) symmetric point as predicted\cite{deppeler,%
azaria_su4}. The DMRG
calculations of Ref. \onlinecite{pati_orbital_dmrg}  showed however a power
law gap opening. Such power law gap opening can only be explained by
the presence of a relevant operator in the continuum description. 
However, no such
operator was obtained in the bosonization treatment\cite{azaria_su4}. 
Moreover, if a
relevant operator was present in the continuum theory, the absence of
a gap at $\lambda/J=4$ would be the result of the coefficient of this
operator vanishing precisely at $\lambda/J=4$. But then,
in contrast to what is observed in numerical calculations, a gap would
also obtain for $\lambda/J>1/4$. A solution to this puzzle has been
suggested recently\cite{yamashita_orbital_dmrg}. 
In Ref.  \onlinecite{pati_orbital_dmrg}, the gap has been 
calculated by assuming that the first excited state was in the subspace
$(S_1,S_2)=(1,1)$. This assumption was shown to be incorrect in the
gapped phase in which the first excited state lies in the subspace
$(S_1,S_2)=(1,0)$. When corrected\cite{yamashita_orbital_dmrg},
a slow increase with
$4-\lambda/J$ is found, compatible with an exponential gap opening. 
The correlation functions \cite{yamashita_orbital_dmrg} do not 
show incommensuration,
in agreement with the field theoretic approach \cite{azaria_su4}.

Taking $J_1=J_1^z \ne J_2=J_2^z$ and $\Delta_1=\Delta_2=1$ in
(\ref{eq:basic_hamiltonian}) preserves the $SU(2)\times SU(2)$
symmetry. This case has been investigated
numerically\cite{yamashita_orbital_dmrg} with $J_1=0.7 J_2-0.3
\lambda/4$. It was shown that a gapless--gapped phase transition
obtained as $J_2/\lambda$ was increased. 
This work was followed by analytical investigation based on the
perturbed $SU(4)_1$ WZW continuum
theory\cite{itoi_spin_orbital,azaria_su4_long}. The analytical
investigations established the existence for a given $\lambda$ 
of an extended gapless region in the plane $J_1--J_2$ that contains
the line $J_1=J_2<\lambda/4$ previously discussed. 

In the present work, we consider the general case of $J_\alpha^z\ne
J_\alpha$ and $\Delta_\alpha \ne 1$. This case includes in particular
the spin tube model. We will focus on the regime $\lambda \ll
J_{1,2}$ and apply methods similar to those of
Ref. \onlinecite{nersesyan_biquad}.

\section{Phase diagram}\label{bosonization}

In this section, we derive the phase diagram of two XXZ spin chains
weakly coupled by a biquadratic exchange. We first recall the
bosonization of a spin chain in Sec. \ref{sec:single_chain}. This
section can be skipped by readers familiar with bosonization. Then, we discuss
the bosonization of the two coupled chains system in
Sec. \ref{sec:biquadratic_coupling}. This allows us to derive
Renormalization Group (RG) equations for the coupling constants of the
problem. Finally, in Sec. \ref{phasediagram}, we discuss the phase
diagram deduced from the analysis of  RG equations.

\subsection {bosonization of a single XXZ chain}\label{sec:single_chain}

In this section, we recall briefly
the derivation of the bosonized Hamiltonian and spin operators that describe
an isolated XXZ chain.
We follow the  well known abelian bosonization procedure for spins
\cite{affleck_houches,solyom_revue_1d,schulz_houches_revue,emery_revue_1d}.
The XXZ spin chain is described by the Hamiltonian:
\begin{equation}\label{eq:hamiltonien_xxz}
H_{\text{XXZ}}=J\sum_i (S_i^x S_{i+1}^x + S_i^y S_{i+1}^y) +J_z \sum_i
S_i^z S_{i+1}^z
\end{equation}
The XXZ spin chain Hamiltonian is first transformed into an interacting
fermionic system on the lattice by expressing the spin operators $S^+$, $S^-$,
$S^z$ in terms of fermion operators $a^{\dagger}$, $a$, using the
Jordan-Wigner transformation\cite{affleck_houches}:

\begin{eqnarray}
\label{jwp}
S_{i}^{+}&=&(-)^i a_{i}^{\dagger}\cos \left (\pi \sum_{j=0}^{i-1}
a^{\dagger}_{ j}a_{ j} \right )\\
\label{jwm}
S_{ i}^{-}&=&(-)^i\cos \left (\pi \sum_{j=0}^{i-1}
a^{\dagger}_{ j}a_{ j} \right )a_{ i}\\
\label{jwz}
S_{ i}^z&=&a_{ i}^{\dagger}a_{ i}-\frac{1}{2}.
\end{eqnarray}

This transformation turns
the XXZ Hamiltonian  into a model of spinless fermions with
nearest neighbor interaction described by the Hamiltonian,
\begin{equation}
\label{spinlessf}
H_{\text{XXZ}}=-\frac{J}{2} \sum_{i,} (a_{ i}^{\dagger}
a_{ i+1}+a_{ i+1}^{\dagger}a_{ i})+J_z \sum_{i,
} (a_{ i}^{\dagger}a_{ i}-\frac 1 2 )
(a_{ i+1}^{\dagger}a_{ i+1}-\frac 1 2 ).
\end{equation}
For $J_z=0$, the Hamiltonian (\ref{spinlessf}) describes
non-interacting fermions and is easily diagonalized. To
proceed, we restrict our attention to the low-energy
sector of the theory, captured by the continuum theory. Introduce
the left (right) chiral fermion fields $\psi_L(x)\;(\psi_R(x))$ 
 containing momenta close to 
the Fermi points $k_F= \pm \frac{\pi } {2a}$, ($x=na$ with
$a$  the lattice spacing),
\begin{equation}
\frac{a_n}{\sqrt{a}}=e^{\imath \frac{\pi n } 2} \psi_R(na) +
e^{-\imath \frac{\pi n} 2} \psi_L(na).
\end{equation}
The continuum Fermi fields are then reexpressed in terms of 
bosonic field\cite{affleck_houches}, as follows:

\begin{eqnarray}\label{eq:bosonized-fermions}
\psi_R(x)=\frac{e^{\imath(\theta-\phi)(x)}} {\sqrt{2\pi a}} \nonumber \\
\psi_L(x)=\frac {e^{\imath(\theta+\phi)(x)}}{\sqrt{2\pi a}} ,
\end{eqnarray}
where
the pair of conjugate fields, $\Pi,\phi$  satisfy the
following commutation relation:
\begin{equation}
[\Phi(x),\Pi(x^\prime)]=\imath \delta(x-x^\prime),
\end{equation}
and the field  $\theta$, dual to $\phi$, is defined as:
\begin{equation}
\theta(x)=\pi \int^x \Pi(x^\prime) dx^\prime.
\end{equation}

 The spin operators Eqs. (\ref{jwp})--(\ref{jwz}), can be
expressed in the continuum limit as:
\begin{eqnarray} \label{spin-operators}
S^+(x)=\frac{S^+_n}{\sqrt{a}}=\frac{e^{\imath
\theta(x)}}{\sqrt{2\pi a}}\left[ e^{\imath \pi \frac x a }+
\cos 2 \phi \right]
\mbox{,} \mbox{ }S^z(x)=\frac{S^z_n}{a}=-\frac {\partial_x \phi} \pi
+\frac{e^{\imath \pi
\frac x a }}{\pi a}
\cos 2 \phi,
\end{eqnarray}

Introducing normal ordering with respect to the fermion vacuum, one has:
\begin{eqnarray}
\label{eq:normal_order}
S_i^x S_{i+1}^x+S_i^y S_{i+1}^y +\Delta S_i^z S_{i+1}^z  & = &  \langle
S_i^x S_{i+1}^x+S_i^y S_{i+1}^y +\Delta S_i^z S_{i+1}^z \rangle + 
\nonumber \\
& + & :S_i^x S_{i+1}^x+S_i^y S_{i+1}^y +\Delta S_i^z S_{i+1}^z :, 
\end{eqnarray}

\noindent where $:\ldots:$ indicates normal ordering. The average $A=\langle
S_i^x S_{i+1}^x+S_i^y S_{i+1}^y +\Delta S_i^z S_{i+1}^z \rangle$
is the ground state energy of the chain and can be found 
 by the Bethe Ansatz\cite{yang_xxz}. However, as long as one is 
only interested in the correlation functions of the single chain, 
one can simply drop this contribution
 and focus on the normal ordered terms
that  describe the excitations above the ground state. 
Our task is thus  to derive a bosonized expression of the normal
ordered product in Eq. (\ref{eq:normal_order}).  
Some care is needed in order to obtain  correct
results\cite{eggert_spin_chains,orignac_2spinchains}, and one finds:
\begin{eqnarray}
\label{eq:bosonized_quad}
:S_i^+S_{i+1}^-+S_i^-S_{i+1}^+: & = & \frac 1 \pi \left[(\pi \Pi)^2 +
(\partial_x \phi)^2\right] + \frac {e^{\frac {\imath \pi x} a}}{\pi}
\sin 2 \phi \nonumber \\
:S_i^zS_{i+1}^z+ S_i^z S_{i+1}^z: & = & \frac 2 {\pi^2} (\partial_x \phi)^2
+ \frac 2 {(\pi a)^2} e^{\imath \frac{\pi x} a} \sin 2 \phi + \frac
{2 \cos 4\phi}{(2\pi a)^2}
+\text{irrelevant terms \ldots}
\end{eqnarray}
The oscillating terms in Eq. (\ref{eq:bosonized_quad}) are dropped  from the
Hamiltonian after integration over $x$.
The Hamiltonian
$H_{\text{XXZ}}$, Eq. (\ref{spinlessf}) then becomes:

\begin{equation}\label{2equivalent-chains-ni}
H_{XXZ}= \int \frac{dx}{2\pi}\left[ u K (\pi \Pi)^2
+\frac u K (\partial_x \phi)^2 \right]-\frac {2\delta}{(2\pi
a)^2} \int dx \cos(4\phi),
\end{equation}
\noindent where for $J_z \ll J$,
\bea
u&=&aJ\left(1+\frac{4J_z}{\pi J}\right)^{1/2}\\
K&=&\left(1+\frac{4J_z}{\pi J}\right)^{-1/2}\\
\delta&=&J_za.\label{deltaequiv}
\eea

 Thus, the bosonized form of $H_{\text{XXZ}}$ reduces
to a  sine-Gordon Hamiltonian,
 where the cosine terms come from the intrachain Umklapp
process\cite{nijs_equivalence,black_equ}.
The renormalization group treatment
shows\cite{shankar_spinless_conductivite}
that in the vicinity of the
XY point, the cosine terms are irrelevant, so that asymptotic
properties are described by  a free scalar field  with renormalized
$u^*,K^*$.
Since the XXZ chain is integrable, it can be shown that the gapless
spectrum extends to the whole region $|J_z|<J$. Moreover, it is
possible to obtain an analytic expression for the renormalized
$u^*,K^*$ from the exact
solution\cite{luther_chaine_xxz,haldane_xxzchain}. One finds:
\begin{equation}
\label{uequiv}
u^*= a \frac {\pi \sqrt {J^2-(J^z)^2}}{2\arccos \frac{J^z}{J}}
\end{equation}
\begin{equation}
\label{kequiv}
K^*= \frac 1 {2-\frac 2 \pi \arccos \frac{J^z}{J}}
\end{equation}
The isotropic point $J_z=J$ (Antiferromagnetic Heisenberg model)
corresponds, in the bosonization description, to $K^*=1/2$ and
$\delta^*=0$. At this point $\cos(4\phi)$ is marginally
irrelevant. For $J_z>J$, the exact solution shows that a gap opens in
the excitations of the
system\cite{haldane_xxzchain,shankar_spinless_conductivite}, and that
an Ising order of the spins
along the z axis  is obtained. This result can also be found from a bosonization
procedure  valid in the
vicinity of the isotropic point.

\subsection{Two coupled XXZ chains with a biquadratic
exchange}\label{sec:biquadratic_coupling}
We now proceed to derive, using the results reviewed in the 
previous subsection, the bosonized Hamiltonian of two
non-equivalent coupled XXZ chains.
To bosonize the Hamiltonian $H_0$, Eq. (\ref{h0}) describing
two decoupled XXZ chain
 we introduce two pairs of dual fields (one pair for each chain)
$\theta_\alpha,\phi_\alpha$ ($\alpha=1,2$)  as defined in
Sec. \ref{sec:single_chain}.
The bosonized form of  the
Hamiltonian $H_0$ is then :

\begin{eqnarray}\label{non-equivalent-chains-ni}
H_0=\sum_{\alpha=1,2} \left\{\int \frac{dx}{2\pi}\left[ u_\alpha K_\alpha
(\pi \Pi_\alpha)^2 +\frac {u_\alpha} {K_\alpha} (\partial_x \phi_\alpha)^2
\right]-\frac{2 \delta_\alpha}{(2\pi a)^2}\int dx \cos 4\phi_\alpha \right\},
\end{eqnarray}
\noindent with the fields $\phi_1$ and $\phi_2$ having \emph{a prori} different
velocities and Luttinger couplings,
\begin{eqnarray}\label{cc-non-equivalent}
& & u_\alpha^* =    \frac {\pi \sqrt {J_\alpha^2-(J_\alpha^z)^2}}{2\arccos
\frac{J_\alpha^z}{J_\alpha}}\nonumber \\
& & K_\alpha^* =  \frac 1 {2-\frac 2 \pi \arccos
\frac{J_\alpha^z}{J_\alpha}} \nonumber \\
& & \delta_{\alpha}=  J_{\alpha}^z a.
\end{eqnarray}
The bosonization formulas
for the spins (\ref{spin-operators}) are unchanged except for the
obvious introduction of a  chain index.

In order to have the full bosonized Hamiltonian, we now have to derive 
the  bosonized form of the biquadratic exchange (\ref{hbiq}).
 The first step is to normal order using Eq. (\ref{eq:normal_order}).
This step is important since it leads to non trivial contributions to the 
quadratic part of the Hamiltonian. 
We introduce the terms:
\begin{eqnarray}
A_\alpha= \lambda \langle
S_{i,\alpha}^x S_{i+1,\alpha}^x+S_{i,\alpha}^y S_{i+1,\alpha}^y
+\Delta_\alpha S_{i,\alpha}^z S_{i+1,\alpha}^z \rangle
\end{eqnarray}
where $\alpha=1,2$ that measure the strength of these corrections. 
Having expressed the Hamiltonian in a normal ordered form, we
can apply  Eq. (\ref{eq:bosonized_quad}) to obtain the bosonized
expression. The oscillating terms of Eq. (\ref{eq:bosonized_quad}) give rise to 
an interchain coupling term, the strength of which is given by:

\begin{equation}
\label{gequiv}
g=2 \lambda a\left(1+2 \Delta_1/\pi\right)\left(1+2 \Delta_2/\pi\right).
\end{equation}
Collecting everything, one obtains the following bosonized biquadratic exchange: 
\begin{eqnarray}
\label{hbiqbos}
H_B& = & \frac {2g}{(2\pi a)^2} \int  dx \sin 2 \phi_1 \sin 2\phi_2 +
 A_1 \int \frac{dx}{2\pi}
\left[(\pi \Pi_2)^2 + (1+2\Delta_2/\pi) (\partial_x \phi_2)^2  \right]
\nonumber \\
&+ & A_2 \int \frac{dx}{\pi}
\left[(\pi \Pi_1)^2 + (1+2\Delta_1/\pi) (\partial_x \phi_1)^2 \right]
+ \text{irrelevant terms \ldots},
\end{eqnarray}
The terms in $A_\alpha$ form a mean-field like interchain interaction .
For $J_{1,2}\ne 0$, these terms  merely produce a renormalization of the
velocities and Luttinger liquid exponents with respect to the
decoupled chains.   
However, in the case where $J_1=0$ or $J_2=0$, these terms become crucial since
they give a finite velocity to the $\phi_1$ (resp. $\phi_2$ excitations)
excitations.  Such a case is realized in the spin-tube problem where
the exchange constant of the pseudospins $\tau$ is zero. We see that
the mean-field like contribution of the spin fluctuations contributes
in that system to provide  
an exchange constant and thus a finite velocity to the pseudospin excitations.
 Of course, in such case, the bosonization procedure is
not really justified since interactions are of the order of magnitude of the
bandwidth of the spin or pseudospin  excitations. However, it is usual in 
quasi-one dimensional systems to have a continuity between the weak and
the strong coupling regime. Moreover, a 
mean field theoretical treatment in the XY limit ($
J_1^z=J_2^z=\Delta_1^z=\Delta_2^z$) leads to similar results to
the bosonization treatment. Details can be found in
App. \ref{app:mean-field}.
We will therefore assume that although not fully justified in the spin
tube case, bosonization nevertheless leads to qualitatively correct
results concerning the phases of the system and the overall behavior
of correlations in these various phases . A quantitative treatment (in
particular of the phase boundaries) requires 
numerical simulations that are beyond the scope of this paper.

The present treatment shows us that in weak coupling the two chains
are described by a bosonized Hamiltonian:
\begin{eqnarray}\label{eq:full_bosonized_hamiltonian}
H & = & \int {\frac {dx}{2\pi}\sum_{\alpha=1,2} \left[ u_\alpha K_\alpha
(\pi \Pi_\alpha)^2 + \frac {u_\alpha}{K_\alpha}(\partial_x
\phi_\alpha)^2 -\frac {4\pi \delta_\alpha}{(2\pi a)^2} \cos
4\phi_\alpha\right]}\nonumber \\
& + &  \frac {2g}{(2 \pi a)^2}\int dx \sin 2\phi_1 \sin 2\phi_2  
\end{eqnarray}

We shall analyze the model perturbatively, with results valid
up to a given value of $\lambda$: it is known  that in the strong
coupling regime of the two chains with biquadratic exchange 
there is a special value of the interchain
coupling at which the model has $SU(4)$ symmetry. At this special
point, the model develops a gapless phase described by a $SU(4)_1$
WZNW model. Such an effect is non-perturbative: the resulting critical
point has a conformal anomaly $c=3$ whereas the original unperturbed
model has $c=2$. This implies by Zamolodchikov's
$c$--theorem\cite{zamolodchikov_c_thm} that the 
transition to the $SU(4)_1$ cannot be predicted by a RG calculation
that  always  leads to a decrease of $c$. 
Beyond this special
value of $\lambda$ , the weak coupling theory would lead to incorrect
predictions and an alternative approach such as the one of Lecheminant
and Azaria is needed. On the other hand, it is expected to give qualitatively
correct predictions when the coupling is smaller than the critical
value. In the remainder of the paper we shall thus work in the weak
coupling regime where the weak coupling theory is valid.
In the following section, we will discuss the RG treatment of the weak
coupling model. 
  
\subsection{Renormalization group equations}\label{sec:rge}

In the following analysis, we will neglect the velocity difference
between chains 1 and 2 since 
usually velocity differences do not play
an important role in the derivation of the phase diagram by RG techniques.
However, for the sake of completeness, we have given in App. \ref{app:momentum_shell_rg} the  RG equations for non-equal
velocities derived using a 
momentum shell integration technique\cite{knops_sine-gordon}.
When velocity differences are neglected, 
 the renormalization group equations  for
 $K_\alpha,\delta_\alpha,g$
  can be easily derived   from the
Hamiltonian (\ref{non-equivalent-chains-ni})--(\ref{hbiqbos}) using
 Operator Product
Expansions (OPE) \cite{cardy_scaling}.
The renormalization group equations for $\delta_\alpha,g$ neglecting
 velocity differences are:
\begin{eqnarray}\label{eq:scaling_rge}
\frac{d}{dl}\left(\frac{\delta_1}{\pi
u}\right)&=&(2-4K_1)\frac{\delta_1}{\pi u}-\frac{g^2}{8\pi^2u^2}
\nonumber \\
\frac{d}{dl}\left(\frac{\delta_2}{\pi
u}\right)&=&(2-4K_2)\frac{\delta_2}{\pi u}-\frac{g^2}{8\pi^2u^2}
\nonumber \\
\frac{d}{dl}\left(\frac{g}{\pi u}\right)&=&(2-K_1-K_2)\frac{g}{\pi
u}-\frac {g(\delta_1+\delta_2)}{2\pi^2 u^2}
\end{eqnarray}
while the renormalization group equations for $K_1,K_2$ are:
\begin{eqnarray}\label{eq:second_order_rge}
\frac {d}{dl}\left(\frac 1 {K_1} \right)=\left(\frac {\delta_1}{\pi
u}\right)^2 + \frac 1 8 \left(\frac g {\pi
u}\right)^2\nonumber \\
\frac {d}{dl}\left(\frac 1 {K_2} \right)=\left(\frac {\delta_2}{\pi
u}\right)^2 + \frac 1 8 \left(\frac g {\pi
u}\right)^2
\end{eqnarray}

\begin{figure}
\centerline{\epsfig{file=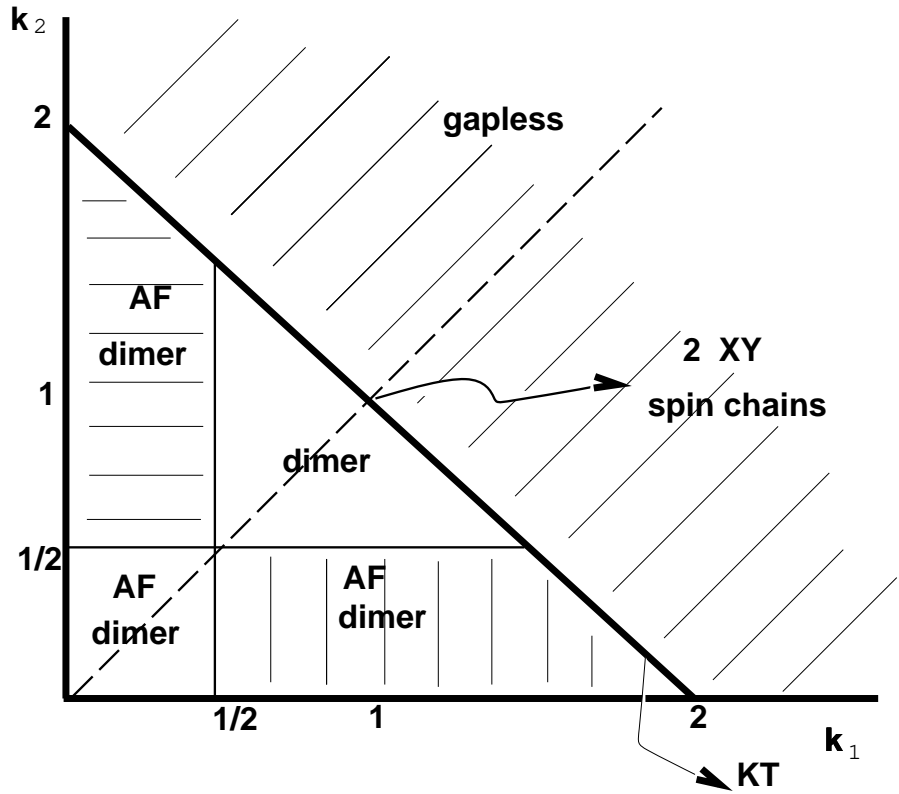,angle=0,width=12cm}}
\vspace{0.5 cm}
\caption{Phase diagram of the spin-tube model. For $K_1+K_2>2$ the
system is in a gapless phase, while it is gapped for $K_1+K_2<2$. The
dashed line corresponds to the isotropic line with $K_1=K_2$. The
isotropic point ($K_1=1$,$K_2=1$) corresponds to two $XY$ chains,
while the isotropic point ($K_1=1/2$,$K_2=1/2$) corresponds to
two-equivalent spin-chains considered by Nersesyan and Tsvelik.}
\label{fig:phase_diagram}
\end{figure}

We now proceed to deduce the weak coupling phase diagram of the model.
  Eq. (\ref{eq:scaling_rge}) indicates that $g$ is a relevant variable when
$(K_1+K_2)<2$, while the variables $\delta_1,\delta_2$ become relevant when
 respectively, $K_1 <1/2$, $K_2<1/2$.
There are therefore four cases to distinguish. In the first case:
$K_1+K_2>2$, there are no relevant operators and the system is in a
gapless state.
In the second case: $K_1+K_2<2$, $K_1,K_2>1/2$, there is a single
relevant operator; for
 $K_1>1/2,\; K_2<1/2$ (or equivalently $K_1<1/2, \; K_2>1/2$), 
 there are  two, while for  $K_1,K_2<1/2$ there are three relevant operators.
These four different regions  are shown in
 Fig.~\ref{fig:phase_diagram}.

In the presence of relevant operators, RG equations cease to be valid
as soon as the dimensionless coupling become $O(1)$, and we have to
determine the nature of the strong coupling fixed points in order to
predict the phase diagram. We see that $K_1,K_2$ are driven to zero 
by the flow of the RG when there is a relevant operator 
so that the fields become classical. As a result, $\phi_1,\phi_2$ are
locked at average values $\langle \phi_1 \rangle,\langle \phi_2
\rangle$  that minimize the ground state energy and a
gap opens in the excitations of these fields\cite{tsvelikb}. When
these fields are locked, it is also known that the exponentials of the dual
field have exponential decay\cite{tsvelikb}.
  
Let us consider  first the case with $g$ the only relevant operator. Then,
the minimization of the ground state energy  requires $\sin 2
\langle \phi_1 \rangle \sin 2 \langle \phi_2 \rangle =-1$,
i.e. $\langle \phi_1 \rangle=-\langle \phi_2 \rangle=\pm \frac \pi
4$. It is then clear that $\cos 2\phi_{1,2}$ as well as $e^{\imath
\theta}$, which are the staggered component of $S^z$ and $S^+$, will
display an exponential decay. It can also be shown that the uniform
component of the spins also present an exponential decay. As a result
we have a spin-liquid phase that presents only short range order in
all its spin correlations.
As we shall see, this spin-liquid phase
is a dimerized phase whose properties  are
 discussed extensively in Sec. \ref{sec:dimerized}. 
In the case $g=0$ (decoupled chains) the analysis is even
simpler. Then the two chains remain decoupled, and we are left with
the analysis of the usual sine-Gordon model. It is then known that
when $\delta_\alpha$  is relevant, the field $\phi_\alpha$ is
locked. The analysis of the resulting strong coupling fixed point can
be found for instance in
Ref. \onlinecite{shankar_spinless_conductivite}. It is found that the
strong coupling fixed point is associated with the Ising
Antiferromagnetic phase of the XXZ chain at $J_z>J$ in which
the staggered component of $S^z$ has a non zero expectation value in
the ground state. 

When  $g$ and at least one of the
$\delta_\alpha$ are relevant, there are  two possible candidates for
the ground state.
Considering Eq. (\ref{eq:scaling_rge}), we see that the effect of the
biquadratic interchain interaction is to reduce the effective
$\delta_{1,2}$. Physically, this means that the tendency to form
singlets competes with the tendency to form an Ising antiferromagnet.
Two  scenarios are possible. One is that there is a well
defined phase boundary between a pure spin-liquid state and a pure
Ising antiferromagnet state. The second scenario is that there is a
crossover between the two pure states as $\lambda/J^z_{1,2}$ is
varied. In such case, increasing $\lambda$ would lead to a gradual
disappearance of antiferromagnetic order leaving a purely singlet state
as $\lambda \to \infty$. Since in both phases it is the same field
that orders, there is \emph{a priori} no reason to exclude a mixed
spin-liquid antiferromagnet order. Thus, the first scenario appears
extremely unlikely. A numerical investigation of the crossover could
be very interesting as a toy model of a crossover from spin-liquid to
antiferromagnetism. 
It is interesting to remark that if
$K_1=K_2=0$, and $\delta_1=\delta_2=\delta$, the equations
(\ref{eq:scaling_rge}) can be integrated analytically. 
Two phases are obtained, separated by a line $g=2\sqrt{2} \delta$. In
the first one, $g \to + \infty$ and $\delta \to -\infty$ which
corresponds to a ground state with singlet order. In  the second one,
$g \to 0$ and $\delta \to +\infty$, which corresponds to
antiferromagnetic order. The RG equations cease to be valid for
$\delta_1/(\pi u) \sim 1$ or $g/(\pi u) \sim 1$. If when this scale is
reached $g$ and $\delta$ are of the same order of magnitude, there is
a possibility of obtaining a mixing of antiferromagnetism and dimer
order. Note that even on the line $g=2\sqrt{2}\delta$, there is a
finite correlation length. This is a further evidence for a
 progressive crossover
from dominant Antiferromagnetic order to dominant dimer order. 

\begin{figure}
\centerline{\epsfig{file=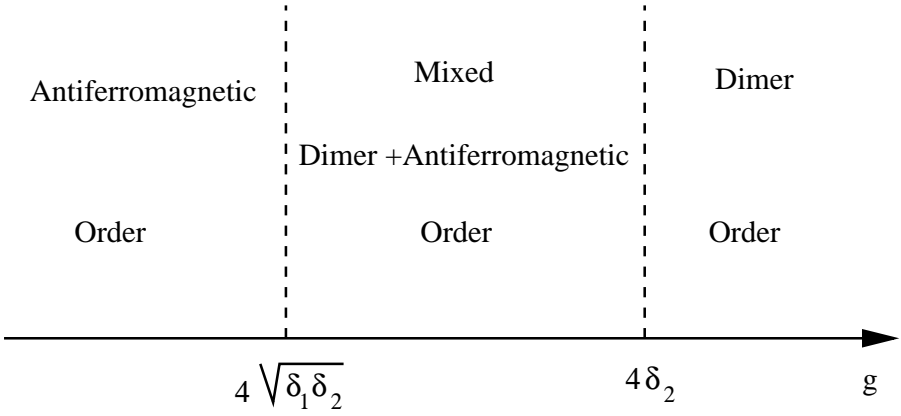,angle=0,width=12cm}}
\vspace{0.5 cm}
\caption{Phase diagram in the limit $K_1=K_2=0$ as a function of $g$
for fixed $\delta_1<\delta_2$. An intermediate phase with dimer order
in chain 1 and mixed  dimer and antiferromagnetic order in chain
2 is obtained when $4\sqrt{\delta_1\delta_2}<g<4\delta_2$.}
\label{fig:classical_phase_diag}
\end{figure}

It is also possible to give a purely classical treatment for $K_1=K_2=0$
by simply minimizing the ground state energy with respect to $\langle
\phi_1 \rangle$ and $\langle \phi_2 \rangle$. In the case
$\delta_2>\delta_1$, one finds that there are three different
regimes. For $g>4\delta_2$ one obtains $\langle \phi_1 \rangle =
-\langle \phi_2 \rangle =\pm \frac \pi 4$ corresponding to a purely
dimerized phase. For $4 \sqrt{\delta_1\delta_2} < g < 4\delta_2$, one
obtains $\langle \phi_1 \rangle =\pm \frac \pi 4$ and $\sin 2 \langle
\phi_2 \rangle =\mp g/(4\delta_2)$. This corresponds to persistence of
dimerization in the chain with the smallest tendency to
antiferromagnetic order, whereas the chain with the strongest tendency
to antiferromagnetism is found in  state with mixed dimer and
antiferromagnetic order. The antiferromagnetic order parameter in that
chain, $\cos 2
\langle \phi_2 \rangle$,  then
assumes the value $\pm \sqrt{1-(g/(4\delta_2))^2}$. Finally in the
region $g < 4\sqrt{\delta_1 \delta_2}$, both chains display
antiferromagnetism with $\langle \phi_{1,2} \rangle = \pm \pi/2$. 
These results are summarized on figure \ref{fig:classical_phase_diag}.  
For $\delta_1=\delta_2$, the region with mixed antiferromagnetic and
dimer order shrinks to a single point. It would be interesting to see
how quantum fluctuations affect the present picture and in particular
determine whether sharp transitions are preserved or if they evolve
into crossovers. 

\subsection{Phase diagram in zero external magnetic field}\label{phasediagram}

In this section, we try to estimate the position of the crossover
between the spin-liquid and the antiferromagnet.
 This can be done roughly by comparing the correlation lengths
 in the antiferromagnet and in the spin-liquid phase.
Using the RG equations, for small $g,\delta_1,\delta_2$, we can
neglect the renormalization of $K_1,K_2$. This leads to:

\begin{eqnarray}
g(l)=g(0)e^{(2-K_1-K_2)l} \nonumber \\
\delta_1(l)=\delta_1(0)e^{(2-4K_1)l} \nonumber \\
\delta_2(l)=\delta_2(0)e^{(2-4K_2)l}
\end{eqnarray}

\noindent when any of these quantities become of the order of the energy cutoff
$\pi v_F/a$, the RG equations cease to be valid and the phase that is
obtained is determined by minimizing the ground state energy.

For $g$, the strong coupling is obtained at a length scale:
\begin{equation}
\label{eq:correlation_dimerized}
L_{\text{dim.}}=a \left(\frac{\pi v_F}{a g}\right)^{\frac 1 {2-K_1-K_2}}
\end{equation}
This is the correlation length of spin fluctuations in the spin-liquid phase.
\noindent For $\delta_1$,$\delta_2$ the strong coupling is obtained respectively at
length scale:
\begin{eqnarray}
L_{\text{AF,1}}=a \left(\frac{\pi v_F}{a |\delta_1|}\right)^{\frac 1
{2-4K_1}} \label{eq:correlation_af_1}  \\
L_{\text{AF,2}}=a \left(\frac{\pi v_F}{a |\delta_2|}\right)^{\frac 1
{2-4K_2}} \label{eq:correlation_af_2}
\end{eqnarray}
It is clear that the shortest length corresponds to the first
operator to attain strong coupling. Therefore, the phase that is
obtained is the one with the shortest correlation length. For
$K_1=K_2=0$, this is in agreement up to a constant with the criterion
derived from the RG equations
(\ref{eq:scaling_rge})--(\ref{eq:second_order_rge}).
The comparison of correlation length allows to draw a rough 
phase boundary between the
antiferromagnet and the dimerized phase that could also
be obtained by numerically integrating the RG equations 
(\ref{eq:scaling_rge})--(\ref{eq:second_order_rge}) starting from
weak coupling and any $K_1,K_2$. 
The equation of the phase boundary is in the case of equivalent chains
$\delta_1=\delta_2=\delta$, $K_1=K_2=K$ :
\begin{equation}
\label{eq:phase_boundary_equivalent}
g=\frac{\pi v_F}{a} \left(\frac{|\delta|a}{\pi
v_F}\right)^{\frac{2-2K}{2-4K}}  
\end{equation}
Let us note that in the isotropic spin-tube case, the operators causing
antiferromagnetic order are (marginally)
 irrelevant so that there is only singlet
order. However, in the case of an anisotropic spin tube (3 coupled XXZ
spin chain with $J_z<J$), such competition becomes
possible. Completely decoupled chains exhibit for $J_z<J$ an Ising 
antiferromagnetic phase. Introducing a strong enough biquadratic
interchain coupling favors on the other hand a spin liquid phase. 
The competition of the two should produce a crossover from the Ising
Antiferromagnet to the spin liquid of the type discussed in the
preceding section.

\section{Spin liquid phase}\label{sec:dimerized}
In this section, we discuss the properties of the spin liquid phase.
 In the case of equivalent chains, further
progress can be made by using symmetric and antisymmetric
modes allowing in particular a
refermionization of the problem and the calculation of some
correlation functions\cite{nersesyan_biquad}. This will be discussed in Sec. \ref{sec:equivalent_chains}. 
In the general case with inequivalent
chains, such decoupling is no longer possible. However, it is still
possible to present a simple semiclassical picture
 of the nature of excitations above the ground state. This will be the 
subject of section \ref{sec:non_equivalent_chains}.

We will focus on
 region $\frac{1}{2}<K_1=K_2<2$, $K_1+K_2<2$, in which the
only relevant operator is the biquadratic exchange,
$\sin2\phi_1\sin2\phi_2$.
The ground state shows
 long-range order of the fields $\phi_1$ and $\phi_2$.
 The expectation values of the ordered
fields are:
\begin{equation}
\langle\phi_1\rangle=\pm\frac{\pi}{4} , \; \langle\phi_2\rangle=\mp\frac{\pi}{4},
\end{equation}
and as a result, $\langle \sin 2 \phi_{1,2} \rangle \ne 0$. 
Using Eq. (\ref{eq:bosonized_quad}), 
this implies that  a dimerized order develops both
in spin variables and pseudospin variables, i.e.
$(-1)^i\langle{\vec S}_{\alpha i} {\vec S}_{\alpha i+1}\rangle \ne 0$
($\alpha=1,2$).
In parallel with that, we have $\langle \cos 2 \phi_{1,2} \rangle= 0$,
so that by Eq. (\ref{spin-operators}) $(-)^i\langle S_{\alpha
i}^z\rangle=0$, and the 
correlation functions $(-)^{|i-j|}\langle S_{\alpha i}^z S_{\alpha j}^z \rangle
\to 0$
as $|i-j|\to \infty$.
It is also well known that when the fields $\phi_a$ are ordered,
 the correlation functions of the disorder operators $e^{\imath p
\theta_{\alpha}}$ decay exponentially at large distances.  Using again
the
bosonization formulas for the spins, Eq. (\ref{spin-operators}), this implies
an exponential decay of all correlation functions :
$(-)^{|i-j|}\langle S_i^a S_j^a \rangle$ and $\langle S_i^a S_j^a
\rangle$ where $a=x,y,z$. Therefore, the dimerized phase appears as a
spin liquid state formed of singlets of spins on both chain 1 and chain 2.
Such a conclusion was reached previously in a numerical investigation of the spin
tube \cite{kawan1} and by considering the equivalent isotropic chains at the
solvable point\cite{mikeska} $\lambda=3J/4$ where the ground state
wavefunction can be obtained exactly in Matrix Product Form. 
Our results show that such mechanism of spin liquid ground state formation
does not require SU(2) symmetry.

This mechanism is somewhat reminiscent of the
spin-Peierls transition\cite{cross_spinpeierls}, the pseudospins
playing here the role of the phonons. This is the essence of the
Mostovoy-Khomskii model\cite{mostovoy} for the ``spin-Peierls'' transition at
$T_c=34$K in $\mathrm{NaV_2O_5}$.

\begin{figure}
\centerline{\epsfig{file=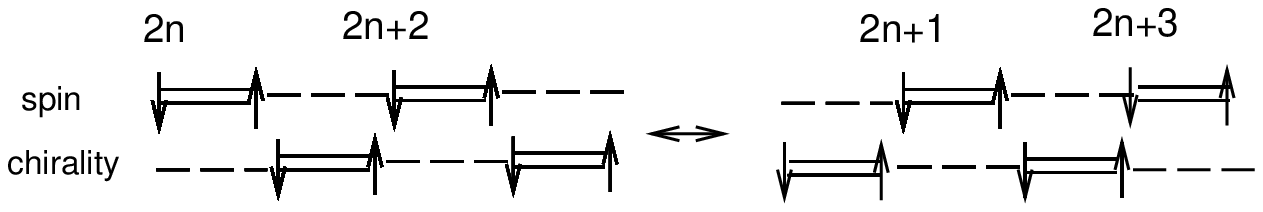,angle=0,width=12cm}}
\vspace{0.5 cm}
\caption{Representation of the two-fold degenerate ground-state,
formed by singlet of spins on even bonds and singlet of chiralities on
odd bonds or singlet of spins on odd bonds and singlet of chiralities
on even bonds.}
\label{fig:ground_state}
\end{figure}

 In the case of the spin-tube the same picture of the ground
state obtains, with  
the ground state of the spin-tube  formed by {\it singlet of spins}
on even bonds and {\it singlet of chiralities} on odd bonds  or by
{\it singlet of spins} on odd bonds and {\it singlet of
chiralities} on even bonds  (see Fig.~\ref{fig:ground_state}).

For the moment, we have only been able to discuss the nature of the
ground state. However, it is also important to  discuss
the nature of the excitations as well as the various correlation functions. 
In order to do that, it is worth to restrict first to the simple case
of two equivalent chains, in which the physical picture is the
clearest. 

\subsection{Equivalent chains}\label{sec:equivalent_chains}

When the two chains are equivalent, we  introduce
the fields\cite{deppeler}:
\begin{equation}
\phi_s=(\phi_1+\phi_2)/\sqrt{2}\mbox{ },\mbox{ }
\phi_a=(\phi_1-\phi_2)/\sqrt{2}
\end{equation}
and their conjugate fields:
\begin{equation}
\Pi_s=(\Pi_1+\Pi_2)/\sqrt{2},\;\; \Pi_a=(\Pi_1-\Pi_2)/\sqrt{2},
\end{equation}

\noindent so that the total Hamiltonian can be completely
decoupled
into the symmetric and antisymmetric parts,

\begin{equation}
H=H_s+H_a,
\end{equation}
\begin{equation}
\label{hs}
H_s=  \int \frac{dx}{2\pi}\left[ u K (\pi \Pi_s)^2
+\frac u K (\partial_x \phi_s)^2 \right] - \frac{g}{(2\pi a)^2} \int dx
\cos \sqrt{8}\phi_s
\end{equation}
\begin{equation}
\label{ha}
H_a=  \int \frac{dx}{2\pi}\left[ u K (\pi \Pi_a)^2
+\frac u K (\partial_x \phi_a)^2 \right] + \frac{g}{(2\pi a)^2} \int dx
\cos \sqrt{8}\phi_a,
\end{equation}

\noindent where the magnetic field couples only to $\phi_s$, and
only the most relevant operators have been taken into
account.
The elementary excitations can be discussed in terms of solitons of
two decoupled sine-Gordon models.
The solitons of the Hamiltonian $H_s$ carry a total
spin $m=\pm 1$ whereas those of Hamiltonian $H_a$  carry a spin 0
as they do not couple to the magnetic field. These
solitons are represented in Fig.~\ref{fig:solitons}.
It is also convenient to use the canonical transformation $\phi_{a,s}
=\tilde \phi_{s,a}/\sqrt{2}$, $\Pi_{s,a} = \sqrt{2}\tilde{\Pi}_{s,a}$
followed by a 
refermionization.
Introducing the fermion operators:
\begin{eqnarray}
\tilde \psi_{R,\nu}(x)=\frac{e^{\imath (\tilde \theta_\nu
-\tilde \phi_\nu)}}{\sqrt{2\pi a}} \nonumber \\
\tilde \psi_{L,\nu}(x)=\frac{e^{\imath (\tilde \theta_\nu
+\tilde \phi_\nu)}}{\sqrt{2\pi a}}
\end{eqnarray}

\noindent where $\nu=a,s$, one can finally rewrite $H_{a,s}$ in the form:
\begin{equation}
H_\nu=-\imath v \int dx (\tilde \psi_{R,\nu}^\dagger \partial_x
\tilde \psi_{R,\nu}
- \tilde \psi_{L,\nu}^\dagger  \partial_x \tilde \psi_{L,\nu}) -\mu_\nu
\int dx
( \tilde \psi_{R,\nu}^\dagger  \tilde \psi_{L,\nu}+ \tilde
\psi_{L,\nu}^\dagger  \tilde \psi_{R,\nu})
+\tilde{\lambda} \int dx \rho_\nu(x)^2
\end{equation}
where $\rho_\nu(x)=\tilde \psi_{L,\nu}^\dagger\tilde
\psi_{L,\nu}+\tilde \psi_{R,\nu}^\dagger  \tilde \psi_{R,\nu}$ and
$\nu=a,s$. The couplings are given by:
\begin{eqnarray}
v &=& 2 u K \nonumber \\
\lambda &=& \pi u \left(\frac 1 {4K} - K\right) \nonumber \\
\frac{g}{4 \pi a} &=& \mu_a=-\mu_s  
\end{eqnarray}

At the isotropic point ($K=1/2$), one has $\lambda=0$, so that $H_\nu$
is a free fermion Hamiltonian
\cite{nersesyan_biquad}.
Similarly to the spin ladder, where $[\vec{S}_{\text{tot.}},H]=0$, the
excitation spectrum can be split into a
singlet and a triplet with spin $-1,0,1$. However, in contrast to the
spin ladder case, the triplet and the singlet here have the same mass. This
is the signature of a larger symmetry group, $SU(2)\times SU(2) \sim
SO(4)$.
The correlation functions can be obtained from mapping the free
fermion Hamiltonian onto two
non-critical Ising model
\cite{haldane_ising,shelton_spin_ladders,nersesyan_biquad} exhibiting the
$SU(2) \times Z_2$ symmetry.
Remarkably, although the system has a spin gap,
these correlation functions are very different from those of a spin-1
chain \cite{tsvelik_field} or a spin ladder
\cite{shelton_spin_ladders}.
In the chain with biquadratic exchange, the 
response functions do not show any particle-like delta function peak
in their imaginary part,
but only a two particle continuum\cite{nersesyan_biquad} even in the
vicinity of $q=\pi$. This is to be contrasted with the spin ladder
\cite{shelton_spin_ladders}
 which shows a delta function peak associated with a single
particle excitation at $q=\pi$. The two chains with biquadratic
interactions thus form a ``non-Haldane'' spin liquid.  

Away from the isotropic point, the Hamiltonian can still be refermionized
but the fermions (solitons) have interactions which preclude a mapping on a
non-critical Ising model. However,  in the
anisotropic case, when the antiferromagnetic intrachain interaction is
irrelevant one has $K>1/2$.
This implies that no bound states of solitons can be
formed since there is no coherent propagation of two
solitons\cite{nersesyan_biquad}, hence the absence of a single particle peak. A calculation of correlation functions is in principle
possible using as form factors approach, but this is far beyond the
scope of the present paper. 
 Since $H_a$ and $H_s$ remain decoupled,
 the excitations having $m=\pm 1$ do not
interact with the excitations having $m=0$. This can be understood as
a consequence of the $U(1)\times U(1)$ symmetry of the problem and the
resulting separate conservation of $S^z_1$ and $S^z_2$.

\subsection{Non equivalent chains}\label{sec:non_equivalent_chains}
In the case of two non-equivalent chains, one cannot decouple the
Hamiltonian into two sine-Gordon Hamiltonians. Therefore, determining
the nature and the quantum numbers 
of elementary excitations is not as straightforward. 
The preceding canonical transformation leads to the following
Hamiltonian: 
 \begin{eqnarray}\label{eq:rotated_non_equivalent}
H& = &\int \frac{dx}{2\pi} \left[ \frac{(u_1K_1+u_2K_2)} 2 [(\pi
\Pi_s)^2+(\pi \Pi_a)^2] + \frac{(u_1/K_1+u_2/K_2)} 2 [(\partial_x
\phi_s)^2 +(\partial_x \phi_a)^2]\right. \nonumber \\
 & + & \left. \frac {u_1 K_1 -u_2 K_2}{2} \pi^2
\Pi_s \Pi_a + \frac{ u_1 /K_1 - u_2 /K_2} 2 \partial_x
\phi_s \partial_x \phi_a \right]+ \frac{2g}{(2\pi a)^2} \int dx [ \cos
\sqrt{8} \phi_s -\cos \sqrt{8} \phi_a ]
\end{eqnarray}
Shifting $\phi_a \to \phi_a+\frac \pi {\sqrt{8}}$, we see that the
resulting Hamiltonian has a $Z_2$ invariance under $ (\Pi_a, \phi_a)
\leftrightarrow (\Pi_s,\phi_s)$. A
rescaling $\phi_a=\tilde \phi_a/\sqrt{2}$, $\phi_s=\tilde
\phi_s/\sqrt{2}$, and a refermionization brings the Hamiltonian to the form: 
\begin{eqnarray}\label{eq:fermion_non_equivalent}
H & = & \int dx \sum_{r=s,a}\left[ -\imath v (\psi^\dagger_{R,r} \partial_x
\psi_{R,r}-\psi^\dagger_{L,r} \partial_x \psi_{L,r}) + m (\psi^\dagger_{R,r} 
\psi_{L,r}-
+\psi^\dagger_{L,r}  \psi_{R,r})\right]   \nonumber \\
& + & 
g(\rho_{R,s}(x)\rho_{R,a}(x)+\rho_{L,s}(x)\rho_{L,a}(x))+ \tilde g_1
\sum_{r\ne r'}\rho_{R,r}(x)\rho_{L,r'}(x) +\tilde g_2 \sum_r \rho_{R,r}(x)\rho_{L,r}(x)
\end{eqnarray}
As a result, there is now an interaction between the excitations of
spin $S^z=0$ (the $a$ fermions) and the excitations of spin $S^z=\pm
1$ (the $s$ fermions). 
We have:
\begin{eqnarray}
v&=&\frac 1 2 \left[ u_1\left(K_1+\frac 1 {4K_1}\right) +
      u_2\left(K_2+\frac 1 {4K_2}\right) \right]\\
g&=&\pi \left[ u_1\left(K_1+\frac 1 {4K_1}\right) -
      u_2\left(K_2+\frac 1 {4K_2}\right)\right] \\
\tilde g_1&=& \pi\left[ u_1\left(\frac 1 {4K_1}-K_1\right) +
      u_2\left(K_2-\frac 1 {4K_2}\right)\right] \\
 \tilde g_2&= &\pi\left[ u_1\left(\frac 1 {4K_1}-K_1\right) -
      u_2\left(K_2-\frac 1 {4K_2}\right)\right] \\
 m& =& \frac{g}{4 \pi a} 
\end{eqnarray}

\begin{figure}
\centerline{\epsfig{file=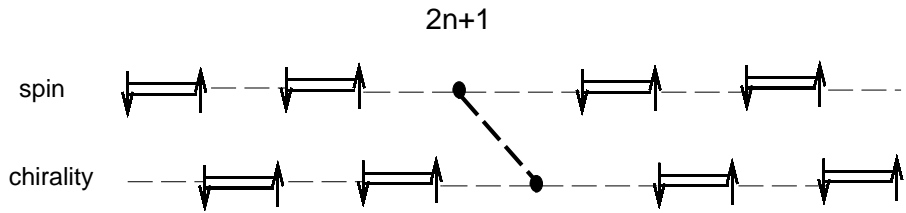,angle=0,width=12cm}}
\vspace{0.5 cm}
\caption{Physical picture of the spin excitations above the ground
state. The elementary excitations confine to form singlets or triplets
between one unpaired spin and one unpaired chirality.}
\label{fig:singlet_triplet_ne}
\end{figure}

The fermionic version of
the model is a generalization of the 
massive Thirring model. The fields carry
spin and the interactions break spin 
rotation symmetry. It can be checked that the interaction of $a$
with $s$ fermions disappears only for $u_1=u_2$, $K_1=K_2$. 
 Not much can be said of the elementary excitations
of Hamiltonian (\ref{eq:fermion_non_equivalent}) due to the absence of
an exact solution.  
In order to gain some insight into the elementary excitations of the
dimerized state in the case of non-equivalent chains, we 
resort to semi-classical approximations. Clearly, we can search for a
semi-classical minimum of (\ref{eq:rotated_non_equivalent}) with
either $\phi_s=0$ or $\phi_a=\pi/\sqrt{8}$. This corresponds to a
single soliton in the system. In the general case, such excitations
are associated with a single fermion, either $a$ or $s$. 
In order to obtain a physical picture for this type of elementary
excitations, let
us  calculate the average  ``magnetization'' for the fields $\phi_1$
and $\phi_2$
 when there is a soliton
connecting the two degenerate dimerized ground-states. In the case
$\phi_s=0$, $\phi_1$ decreases from  $\pi/4$ to $-\pi/4$ while
$\phi_2$ increases from  $-\pi/4$ to $\pi/4$. We immediately get:

\begin{eqnarray}
\label{ms}
m_1 &=&-\frac{1}{\pi}\int_{-\infty}^{\infty}\partial_x
\phi_1=-\frac{1}{\pi}\lbrack
\phi_1(+\infty)-\phi_1(-\infty)\rbrack=-\frac{1}{2} \\
\label{mt}
m_2&=&-\frac{1}{\pi}\int_{-\infty}^{\infty}\partial_x
\phi_2=-\frac{1}{\pi}\lbrack
\phi_2(+\infty)-\phi_2(-\infty)\rbrack=-m_1=\frac{1}{2}.
\end{eqnarray}
Another solution is obtained by reversing the sign of $\phi_a$ leading
to $m_1=-m_2=1/2$. 
The case $\phi_a=\pi/\sqrt{8}$ corresponds to $m_1=m_2=\pm 1/2$. In
each case, the elementary excitations are formed by  breaking a singlet
on  neighboring sites on each chain. Such objects then propagate
coherently.  
\noindent In the case of the spin tube, this
 corresponds to having  one unpaired spin
associated with one unpaired chirality pseudospin. Such an excitation
has $\tau^z=\pm 1/2$, $S^z=\pm 1/2$. It is the spinon of
Ref. \onlinecite{cabra_unp}. 
Thus the elementary excitations of the
model can be easily visualized as  an unpaired spin and an
unpaired chirality forming a triplet or a singlet diagonal bond
(see Fig.~\ref{fig:singlet_triplet_ne}). This is the soliton in the
dimer order.
This picture generalizes from isotropic 
equivalent spin chains \cite{nersesyan_biquad} to
the case of inequivalent and anisotropic  chains. An open question
is whether in the case of inequivalent chains bound states of these
elementary excitations can be obtained in contrast to the case of
equivalent chains. A necessary condition is that there exists
attractive interactions between $a$ and $s$ fermions. The study of
such bound states could be of interest in relation with light
scattering experiments on $\mathrm{NaV_2O_5}$.  
Besides semi-classical approximation, another approximate treatment is
possible. In the case where $u_1=u_2$, $\delta_1=\delta_2=0$ and
$K_1+K_2=1$, the Hamiltonian (\ref{eq:full_bosonized_hamiltonian}) is
 the double sine Gordon
model\cite{fateev_family_integrable,lesage_qwires_ir_fp}, or
the Bukhvostov-Lipatov
model\cite{bukhvostov_lipatov}. 
The model is known to be Bethe Ansatz
solvable\cite{bukhvostov_lipatov} but its elementary excitations and
$S$ matrix have
only been obtained recently \cite{saleur_bl_model} and shown to be
identical to those of the double sine-Gordon model on its integrable
line. The spectrum of the
 model \cite{saleur_bl_model} is consists  of four massive particles carrying
two quantum numbers $Q_1,Q_2=\pm 1$, and the $S$ matrix 
is\cite{fateev_family_integrable,lesage_qwires_ir_fp,saleur_bl_model}: 
\begin{equation}
\label{eq:smatrix_bl}
S=S_{\text{SG}}^{\hat{K_1}}\otimes S_{\text{SG}}^{\hat{K_2}},
\end{equation}
where $S_{\text{SG}}^K$ is the $S$ matrix of a sine Gordon model
having the parameter $K$ and $\hat{K_\alpha}=4K_\alpha/(1+2K_\alpha)$.
As a result, one of the sine Gordon models is in the attractive regime
and has a spectrum made of fermions and bound states of fermions
whereas the second sine-Gordon model is is the repulsive regime and
has a spectrum containing only fermions. The mass of the fundamental
fermion is $m=\frac{g}{4\pi a \sin(\pi K_1)}$. For $K_1=1/2$, one
recovers the equivalent chain and the mass of the bound states is then:
\begin{equation}
m_n=\frac{g}{4\pi a} \sin(\pi n K_1) , n \in N, nK_1< 1/2
\end{equation}
 
Since for $1/4<K_1<3/4$, the operator $\sin 2\phi_1 \sin 2\phi_2$ is
the most relevant, it is reasonable to expect that neither the marginal 
perturbations due to $u_1\ne u_2$ nor the less relevant
perturbations $\delta_{1,2}\ne 0$ change the gapped nature of the
spectrum. In this regime, one should not observe any bound state,
since the $n=2$ bound state only exists for $K<1/4$. In terms of the
original spin chain, the double sine-Gordon regime should be
accessible if one chooses to have one chain with $J_z<J$ and the other
chain with $J_z>J$ in such way that $K_1+K_2=1$ and $J_\perp \ll
J,J_z$. However, the double sine-Gordon regime would not be observed
in the spin tube case, in which we should have $K_1+K_2=3/2$. We
expect that in the spin tube case, the system has more quantum
fluctuations than in the double sine-Gordon regime. Therefore, no
bound states of spinons will form. This heuristic argument agrees
with the numerical calculations on the spin tube that show no 
bound state of spinons\cite{kawan1,cabra_unp}.

\section{Effect of a magnetic field on the dimerized
phase}\label{sec:gap_closure}

In this section, we discuss the effect of the application of 
a magnetic field on the
dimerized phase. In general, the application of a magnetic field to a
gapped one dimensional spin liquid system results in the closure of
the gap for a magnetic field of the order of magnitude of the gap. 
Below the critical magnetic field the magnetization is zero. Above the
critical magnetic field the magnetization increases as
$(h-h_c)^{1/2}$, the system is in a single component Luttinger liquid
state and has incommensurate spin
correlations\cite{chitra_spinchains_field}. 
Here, we  have to consider two \emph{a priori} different cases. In the
first one and most academic,
the two coupled chains carry real spins that couple to the magnetic
field. In the second case, only one
chain carry a spin that couple to the magnetic field, 
the other one carrying only a pseudospin degree of
freedom that does not couple to a magnetic field. This last case
corresponds to the spin-orbital models\cite{mostovoy} and to the spin
tube model\cite{schulz_moriond}. The behavior of the spin-orbital
model in a magnetic field has been discussed in
Refs. \onlinecite{azaria_su4_long} and
\onlinecite{lee_spin_orbital_mf}. Both references study models
with  $SU(2) \times SU(2)$ symmetry in the absence of the magnetic
field. It is shown that in the vicinity of the SU(4) point, the
spin-orbital model becomes equivalent to the $O(4)$ Gross Neveu model
describing the gapped modes 
plus a $c=1$ CFT that describes the gapless magnetic modes. Also, in
Ref. \onlinecite{lee_spin_orbital_mf}, the weak coupling case has been 
discussed.   
The  case where both chains carry spin will be dealt with in
Sec. \ref{sec:both_w_spin} and the case of a single chain carrying
spin in 
Sec. \ref{sec:one_w_spin}.

\subsection{Both chains carry spin}\label{sec:both_w_spin}
If both chains carry spin, the coupling to the magnetic field (taken
parallel to the z axis) is:
\begin{equation}
-\frac{h}{\pi}\int dx \partial_x(\phi_1 + \phi_2)
\end{equation}
 In that case, we can use the decoupling of
Sec. \ref{sec:equivalent_chains} , Eqs. (\ref{hs}) and (\ref{ha}).
The coupling to the magnetic field is:
\begin{equation}
-\sqrt{2} \frac h \pi \int dx \partial_x \phi_s
\end{equation}
\subsubsection{The case of equivalent chains}
We consider the case where the two chains are equivalent.
This problem has  been discussed 
 in the context of spin ladders\cite{cithra}. 
The gap in
the antisymmetric modes $\phi_a$
is not affected by the presence of the magnetic field. On the other
hand, for a sufficiently large magnetic field $h>h_{c_1}$, the gap in the
symmetric modes closes and the magnetization $m=-\frac {\sqrt{2}}{  \pi}
\partial_x \phi_s$ behaves as $m \sim (h-h_{c_1})^{1/2}$ for
 $h>h_{c_1}$. Moreover, it can be shown that the low energy modes of
 $\phi_s$ are described by the following Hamiltonian:
\begin{equation}
H_s=\int \frac{dx}{2\pi}\left[u_s^* K_s^* (\pi \Pi_s)^2 +\frac{u_s^*}{K_s^*}
(\partial_x \phi_s)^2 \right]
\end{equation}
And the exponent $K_s$ takes the universal value 1/2 at $h=h_c$. 
Correlation functions in the incommensurate phase can be obtained in
 the isotropic case by a
 calculation similar to the spin-ladder case
 \cite{furusaki_dynamical_ladder}. This time, $\langle \phi_a^2 \rangle$
 is finite so that $e^{\imath \theta_a/\sqrt{2}}$ has exponentially
 decaying correlations. As a result, one has:
\begin{equation}
\langle S^+(x) S^-(0) \rangle \sim e^{-x/\xi}, x \ll \xi
\end{equation}
A simplified expression for the operator $S^z=S_1^z+S_2^z$ is:
\begin{equation}
S^z=-\frac{\sqrt{2}}{\pi}\partial_x \phi_s + \cos (\sqrt{8} \phi_s
-2\pi m x) 
\end{equation}
and the correlations of $S^z$ are:
\begin{equation}
\langle T_\tau S^z(x,\tau) S^z(0,0)\rangle = \frac{x^2-(u
\tau)^2}{(x^2+(u\tau)^2)^2} + \text{constant}\left(\frac
{a^2}{x^2+(u\tau)^2}\right)^{2K_s^*}  \cos (2 \pi m x)
\end{equation}
There are subleading power law
corrections at $Q=2 \pi m x$. As shown by Furusaki and Zhang, these
corrections are missed if one naively neglects the band curvature
after refermionization \cite{cithra}. These corrections can also be
obtained\cite{giamarchi_coupled_ladders} by using the Haldane
expansion of the spin operators and retaining the terms up to $4k_F$
as we did here. 
\subsubsection{The case of non-equivalent chains}
In the case of non-equivalent chains, the problem gets more
complicated. The magnetic field still couples only to $\phi_s$. 
However, when the magnetic field $h$ exceeds the field $h_{c_1}$
needed to close the gap in the $s$ modes, 
the appearance of a non-zero $\langle \partial_x \phi_s\rangle$
creates an effective magnetic field that couples to $\partial_x
\phi_a$. If this magnetic field is not strong enough to close the gap
in the $\phi_a$ modes, the system remains in a one-component
Luttinger liquid for fields $h$ not much stronger than the critical
field $h_{c_1}$. For $h$ sufficiently large, the generated effective
field can close the gap in the $a$ modes leading to a two component
Luttinger Liquid. Since no experimental ladder system with a biquadratic
exchange much larger than the quadratic exchange much larger that the
quadratic exchange and made of two non-equivalent chains is presently
available, this two-step transition to a two component Luttinger
liquid is unlikely to be observed experimentally.   
 
\subsection{Only one chain carries spin}\label{sec:one_w_spin}
This case includes the spin tube and the Mostovoy-Khomskii
model and is therefore the most relevant physically. This case has
been discussed for two coupled $SU(2)\times SU(2)$ symmetric chains in
Ref. \onlinecite{lee_spin_orbital_mf}. 
In the case where only one chain, say chain $1$ to fix notations,
carries spin, the interaction with the magnetic field is given by a
term: 
\begin{equation}
-\frac{h}{\pi}\int dx \partial_x \phi_1,
\end{equation}
 In
the case of equivalent chains, we can use the same decoupling of
Sec. \ref{sec:equivalent_chains} , Eqs. (\ref{hs}) and (\ref{ha}) as
in the preceding section. However, there is an important difference.
Now, we have the coupling to the magnetic field in the form:
\begin{equation}
-\frac h {\pi \sqrt{2}} \int dx \partial_x (\phi_s +\phi_a)
\end{equation}
As a result, now the magnetic field couples to both
\cite{lee_spin_orbital_mf}  $\phi_s$
and $\phi_a$. Moreover, the strength of the couplings is exactly the
same. We start with the discussion of equivalent chains. Then, we
discuss non equivalent chain. We will show that in both case, the
closure of the gap leads to a two-component Luttinger liquid behavior
in contrast with the usual spin-liquid systems that lead to a single
component Luttinger liquid behavior\cite{chitra_spinchains_field}.
We will also give the expression of correlation functions in the
Luttinger liquid phase.

\subsubsection{Equivalent chains}
 As a result of the symmetry between $\phi_a$ and $\phi_s$, 
the gap in the symmetric and the antisymmetric mode
close simultaneously, leading to a two component Luttinger liquid
ground state under strong enough magnetic field. Contrarily to the
case where the magnetic field couples to both 1 and 2 spins, there is
no intermediate single component Luttinger liquid phase. 
We expect that the
magnetization $m=-\partial_x \phi_1/\pi$ behaves as $(h-h_{c_1})^{1/2}$
close to the threshold.
It is also important to note that, the two sine-Gordon model being
equivalent, one has for the fixed point Hamiltonian $u_a^*=u_s^*$ and
$K_a^*=K_s^*$. As a result, the fixed point Hamiltonian is invariant
under any rotation in the $(\phi_a,\phi_s)$ plane.
We can thus write the fixed point Hamiltonian as:
\begin{equation}\label{eq:fp_mf_equiv}
H=\int \frac{dx}{2\pi}\left[u^* K^* (\pi\vec{\Pi})^2 +
\frac{u^*}{K^*}(\partial_x \vec \phi)^2\right]
\end{equation}
Where $\vec \phi=(\phi_1,\phi_2)$ and $\vec \Pi=(\Pi_1,\Pi_2)$.
Moreover, in the case of $SU(2) \times SU(2)$ symmetry, we have
$K_1=K_2=1/2$. It can the be shown easily, using the refermionization
procedure  that  $K^*=1/2$ for any magnetic field.
Equation (\ref{eq:fp_mf_equiv})
has important consequences for the correlation functions, which
are of the form (for equivalent chains):

\begin{eqnarray}
\langle S^+_{\alpha}(x,t) S^-_{\alpha}(0,0)\rangle  = 
(-1)^x(x^2-t^2)^{-\frac{1}{4K_\alpha}} + \text{constant} \mbox{ }
\left[   \frac{e^{2i\pi m_\alpha
x}}{(x-t)^{2(K_\alpha+\frac{1}{4K_\alpha})}} + \frac{e^{2i\pi m_\alpha
x}}{(x+t)^{2(K_\alpha+\frac{1}{4K_\alpha})}} 
\right]  \nonumber \\
\langle (S^z_{\alpha}(x,t)-m_\alpha)
( S^z_{\alpha}(0,0)-m_\alpha)
\rangle  = 
 \cos(\pi x (1-2m_\alpha))(x^2-t^2)^{-K_\alpha}+ \text{constant} \mbox{
}\frac{1}{4\pi^2}
\left ( \frac{x^2+t^2}{(x^2-t^2)^2} \right ),
\nonumber 
\end{eqnarray}

\noindent where the index $\alpha$ indicates the spin in chain 1 or 2,
$m_\alpha$ is the magnetization, with $m_1=m$ (total magnetization)
 and $m_2=0$ . From the above
expressions we deduce the following: the correlation function parallel to
the field, $\langle S^z_1 S^z_1 \rangle$, has a staggered part shifted
from the wave vector $q=\pi$ to $q=\pi(1 -2 m)$, while the correlation
function perpendicular to the field, $<S^+_1S^{-}_1>$, has
an unshifted staggered mode and  the uniform magnetization mode
shifted to $q=2\pi m$. The correlation functions for the spin of type
2, instead, are completely unaffected by the presence of an external
magnetic field. 
\subsubsection{General case}
In this section, we consider the case where the two chains are not
necessarily equivalent. In particular, this is the case that is
realized in the spin-tube under a magnetic field.

In this case, is convenient to use the rotation,
\begin{eqnarray}
\phi_s&=&(\phi_1+\phi_2)/\sqrt{2},\; \; \;
\phi_a=(\phi_1-\phi_2)/\sqrt{2}\nonumber \\
\Pi_s&=&(\Pi_1+\Pi_2)/\sqrt{2}, \;\Pi_a=(\Pi_1-\Pi_2)/\sqrt{2}
\end{eqnarray}
to bring  the Hamiltonian to the form:
\begin{eqnarray}\label{eq:rotated_mf}
H& = &\int \frac{dx}{2\pi} \left[ \frac{(u_1K_1+u_2K_2)} 2 [(\pi
\Pi_s)^2+(\pi \Pi_a)^2] + \frac{(u_1/K_1+u_2/K_2)} 2 [(\partial_x
\phi_s)^2 +(\partial_x \phi_a)^2]\right. \nonumber \\
 & + & \left. \frac {u_1 K_1 -u_2 K_2}{2} \pi^2
\Pi_s \Pi_a + \frac{ u_1 /K_1 - u_2 /K_2} 2 \partial_x
\phi_s \partial_x \phi_a \right]+\nonumber \\
& + & \frac{2g}{(2\pi a)^2} \int dx [ \cos
\sqrt{8} \phi_s -\cos \sqrt{8} \phi_a ] -\frac h {\sqrt{2}} \int dx
\partial_x(\phi_s+\phi_a)
\end{eqnarray}

Shifting $\phi_a \to \phi_a +\pi/\sqrt{8}$, renders
Hamiltonian  invariant under the interchange of $\phi_s$ and
$\phi_a$. As a consequence, the gaps in $\phi_s$ and $\phi_a$ are
\emph{identical} and have to close simultaneously.
Therefore, as in the case of identical chain, there is a transition
from a gapped phase into a gapless two-component Luttinger Liquid
phase. The most general Hamiltonian for a two-component Luttinger
liquid is:
\begin{eqnarray}
\label{eq:2component_ll}
H& = & \int \frac{dx}{2\pi} \left[M_{aa}(\pi \Pi_a)^2 + 2 M_{as} \pi \Pi_a \pi
\Pi_s + M_{ss} (\pi \Pi_s)^2 + \right . \nonumber \\
 & & \left . +  N_{aa} (\partial_x \phi_a)^2 + 2N_{as}
\partial_x \phi_a \partial_x \phi_s +N_{ss} (\partial_x \phi_s)^2 \right].
\end{eqnarray}
In the case of coupled XXZ chain, the effective Hamiltonian
(\ref{eq:2component_ll})  can be
simplified by making use of symmetry.
We know that the original Hamiltonian, Eq. (\ref{eq:rotated_mf}) is
invariant under the transformation:
\begin{eqnarray}
\phi_a &\leftrightarrow & \phi_s \nonumber \\
\Pi_a &\leftrightarrow & \Pi_s
\end{eqnarray}
This symmetry must be preserved by the renormalized
Hamiltonian. Therefore, one must have $M_{aa}=M_{ss}$ and
$N_{aa}=N_{ss}$. As a result, the effective Hamiltonian
(\ref{eq:2component_ll}) is
diagonalized by returning to the original variables
$\phi_1,\; \phi_2$.
The effective Hamiltonian is therefore the following:
\begin{equation}\label{eq:effective_gapless}
H=\int \frac{dx}{2\pi} \left[u_1^*K_1^* (\pi \Pi_1)^2 +
\frac{u_1^*}{K_1^*} (\partial_x \phi_1)^2 + u_2^* K_2^* (\pi \Pi_2)^2
+ \frac{u_2^*}{K_2^*} (\partial_x \phi_2)^2 \right]
\end{equation}
The absence of coupling between $\phi_1$ and $\phi_2$ in this
Hamiltonian is somewhat surprising. This can however be understood by
the fact that the Hamiltonian even in the presence of a magnetic field
is invariant by a rotation by $\pi$ $S_2^{y,z}\to -S_2^{y,z}$. Thus,
$\langle S^z_2\rangle=0$ for any $h$, implying that $\partial_x \phi_1
\partial_x \phi_2$ terms cannot appear in the two component Luttinger
liquid Hamiltonian.
In contrast to the spin ladder case\cite{cithra}, the exponents $K_{1,2}^*$
in the Hamiltonian Eq. (\ref{eq:effective_gapless}) are
\emph{non-universal}. This can be shown in the following way. Let us
consider the case where $u_1K_1 - u_2 K_2$ and $u_1/K_1 - u_2/K_2$ are
small compared to $u_1 + u_2 $, namely where the couplings in
the two chains are
nearly identical. In that case, we can neglect in a
first approximation the terms $\Pi_a \Pi_b$ and $\partial_x \phi_a
\partial_x \phi_b$ in the Hamiltonian \ref{eq:rotated_mf}. We are thus
left with two decoupled sine-Gordon models under a magnetic field
. These sine-Gordon models undergo a commensurate-incommensurate
transition at a critical magnetic field.
It is well known\cite{schulz_cic2d} that the exponent at the
 transition assumes a universal value that
renders scaling dimension of the cosine term equal to one. Therefore, in
our case, the universal value of the exponent at the transition is:
$K^*=1/2$.
As a result, close to the transition the effective Hamiltonian is:
\begin{eqnarray}
H=\int \frac{dx}{2\pi} \sum_{\nu=a,s} \left[u^* K^* (\pi \Pi_\nu)^2 +
\frac{u^*}{K^*}(\partial_x \phi_\nu)^2\right] +  \int \frac{dx}{2\pi}
\left[ \frac {u_1 K_1 -u_2 K_2}{2} \pi^2
\Pi_s \Pi_a +  \frac{ u_1 /K_1 - u_2 /K_2} 2 \partial_x
\phi_s \partial_x \phi_a \right].
\end{eqnarray}
Returning to $\phi_1$ and $\phi_2$, one obtains an Hamiltonian of the
form Eq. (\ref{eq:effective_gapless}) with:
\begin{eqnarray}
K_1^*&=&\sqrt{\frac{1 +\left(\frac{u_1K_1-u_2K_2}{u^*}\right)}{1
+\left(\frac{u_1K_1}{u^*}-\frac{u_2K_2}{u^*}\right)}} \nonumber \\
u_1^*&=&u^*\sqrt{\left(1 +\left(\frac{u_1K_1-u_2K_2}{u^*}\right)\right)\left(1
+\left(\frac{u_1K_1}{u^*}-\frac{u_2K_2}{u^*}\right)\right)}\nonumber
\\
K_2^*&=&\sqrt{\frac{1 -\left(\frac{u_1K_1-u_2K_2}{u^*}\right)}{1
-\left(\frac{u_1K_1}{u^*}-\frac{u_2K_2}{u^*}\right)}} \nonumber \\
u_2^*&=&u^*\sqrt{\left(1 -\left(\frac{u_1K_1-u_2K_2}{u^*}\right)\right)\left(1
-\left(\frac{u_1K_1}{u^*}-\frac{u_2K_2}{u^*}\right)\right)}
\end{eqnarray}
indicating that except in the case of equivalent chains, one
should not expect universal exponents at the transition. This should
be the case in particular for the spin-tube \cite{schulz_moriond}.
It may provide an experimental test for the
 spin-orbital model\cite{mostovoy,pati_orbital_dmrg} of
$\mathrm{NaV_2O_5}$
since the exponent $K_1^*$ controls the temperature dependence of the
NMR relaxation rate\cite{sachdev_nmr,cithra}. However, since the
transition temperature between the gapped and the gapless phase
in $\mathrm{NaV_2O_5}$ is $T_c= 35 K$, the magnetic field needed to
close the gap should be of order $52.5 T$ which could make the
experiment impossible. In $\mathrm{Na_2 Ti_2 Sb_2 O}$, with
$T_c=110K$, the situation is even worse.

The spin and pseudospin operators are,
\begin{eqnarray}
S^z(x)&=& m-\frac{\partial_x \phi_1} {\pi} +\frac{e^{\imath \frac{\pi x}
a}}{\pi a}
\sin (2\phi_1- 2\pi m x),
\mbox{ } \mbox{ } 
S^+(x)=\frac{e^{\imath \theta_1}}{\sqrt{\pi a}} \left[ e^{\imath
{\pi x} a} + \sin (2\phi_1 - 2 \pi m x) \right] \nonumber \\
\tau^z(x)&=&-\frac{\partial_x \phi_2} {\pi} +
\frac{e^{\imath \frac{\pi x} a}}{\pi a}
\sin 2\phi_2,
\mbox{ } \mbox{ } \;\;\;\;\;\;\;\;\;\;\;\;\;\;
\tau^+(x)=\frac{e^{\imath \theta_2}}{\sqrt{\pi a}} \left[ e^{\imath
{\pi x} a} + \sin 2\phi_2  \right],
\end{eqnarray}
\noindent where  $m$ is the total magnetization.
In the case of the spin-orbital model, the spin-spin correlation
functions are therefore given by the usual
formulas\cite{affleck_houches}.
The situation is however more interesting in the case of the spin
tube. Using the formula (\ref{eq:spin_stot_chirality}), one has:
\begin{equation}
\langle T S_p^\beta(x,t) S_p^\beta(0,0)\rangle = \langle
T S^\beta(x,t)S^\beta(0,0)\rangle \times \left[\frac 1 9
+ \frac 8 9 \langle T \tau^+(x,t) \tau^-(0,0)
\rangle\right],
\end{equation}
\noindent where $\beta=(+,-,z)$, and $p$ is the chain index. Explicitly,
\begin{eqnarray}
\langle T S^+(x,t) S^-(0,0)\rangle =
(-1)^{x/a} (x^2-(u_1 t)^2)^{-\frac{1}{4K_1}} +  \text{const} \mbox{ }
(x^2-(u_1 t)^2)^{-(\frac{1}{4K_1} +K_1 -1)}
\times \lbrack \frac{e^{2i\pi m x}}{(x - u_1 t)^2} +
\frac{e^{-2i\pi m x}}{(x + u_1 t)^2} \rbrack\nonumber \\
 \langle (S^z(x,t)-m)
( S^z(0,0)-m)\rangle = 
 \cos(\pi x (1-2m))(x^2-(u_1t)^2)^{-K_1}+ \text{const} \mbox{ }
\frac{K_1}{4\pi^2}
\left ( \frac{1}{(x-u_1 t)^2} +\frac{1}{(x+ u_1 t)^2}\right )
\nonumber \\
 \langle T \tau^+(x,t) \tau^-(0,0)\rangle =
(-1)^x(x^2-(u_2 t)^2)^{-\frac{1}{4K_2}} + \text{const} \mbox{ }
(x^2-t^2)^{-(\frac{1}{4K_2} + K_2 -1)}
\left (  \frac{2(x^2+(u_2 t)^2)}{(x^2-(u_2 t)^2)^2}\right ).\nonumber
\end{eqnarray}

This correlation function, which enters in particular
in the calculation of the NMR relaxation rate, contains power-law
divergences at wave vector $q\sim 0, \pi(1\pm 2m)$ but also $\pm 2m$
due to the fluctuations of chiralities.

\section{Conclusions}\label{conclusions}

We have presented a field-theoretical analysis of 
 the low-energy physics  of the 
anisotropic spin-orbital model and the three-leg
ladder with periodic boundary conditions  (the spin tube) in the strong
 interchain coupling limit. We gave a derivation of the field
 theoretical model from the lattice Hamiltonian and then analyzed the
phase-diagram  using renormalization
group equations. The system is found to exhibit a
 gapless phase, a spin-liquid phase or an Ising Antiferromagnetic
 phase depending on the microscopic couplings. The spin liquid ground
 state  is two-fold degenerate,
formed either by singlets of spins
on even bonds and singlets of orbital pseudospins (chirality in the
 spin-tube case) on odd bonds or the other way round.
 The antiferromagnetic
 phase competes with the spin-liquid and we have discussed this
 competition briefly. The spin liquid phase  obtains in the
 case of the spin tube with $SU(2)$ symmetry, and we
  discussed the nature of excitations above this
 spin liquid ground state.
 These excitations have spin $S^z=\pm 1 2$ and pseudospin 
 $\tau^z =\pm 1 2$. They are formed by introducing a free spin as a defect in
 the spin singlet pattern as well as a pseudospin in the pseudospin singlet
 pattern. These excitations lead to a kind of non Haldane spin liquid 
analogous to the one discussed by Nersesyan and
 Tsvelik\cite{nersesyan_biquad}. An interesting consequence is the
 absence of a magnon peak at $q=\frac \pi a$ in the spin-spin
 correlation functions of the spin tube. This behavior could be tested
 in numerical simulations.
We have investigated the effect of an applied magnetic field $h$ on the dimerized
phase.  A strong enough magnetic field $h>h_{c_1}$
 causes
the closure of the gap and the disappearance of dimer order. 
 The resulting gapless phase is a \emph{two} component
Luttinger liquid in contrast to the one component Luttinger Liquid
that is observed in spin ladders. The exponents appeared to be
 \emph{non-universal} at the transition point, in contrast with the
 spin ladder case\cite{chitra_spinchains_field}. 
It would be interesting to obtain numerically the Luttinger liquid
 exponents for the spin tube or the anisotropic spin orbital model.
For the spin tube, we have also shown that new soft modes appeared in
 the spin-spin correlation functions above $h_{c_1}$ by comparison
 with the soft modes of the single chain. This is the result of the
 presence of soft chirality modes. This may be tested numerically.

\acknowledgments
E. O. acknowledges support from NSF under grant NSF-DMR 96-14999.
We thank P. Lecheminant and H. Saleur
 for discussions and useful comments on the
manuscript.

\noindent $^\dagger$ {\footnotesize Permanent address:  CNRS-Laboratoire de
Physique Th\'eorique de L'Ecole Normale Sup\'erieure 24, Rue Lhomond 75231 Paris Cedex
05 France}
\newpage

\appendix

\section{Mean Field treatment of the two XY chains coupled with a
biquadratic exchange respecting the XY symmetry}\label{app:mean-field}

\subsection{Equivalent chains}
We start from the Hamiltonian:
\begin{equation}
H=\sum_{n \atop a=1,2} S_{n,a}^+S_{n+1,a}^- + S_{n,a}^-S_{n+1,a}^+ +
\lambda \sum_n (S_{n,2}^+S_{n+1,2}^- +
S_{n,2}^-S_{n+1,2}^+)(S_{n,1}^+S_{n+1,1}^- + S_{n,1}^-S_{n+1,1}^+)
\end{equation}
After the usual Jordan Wigner Fermionization, this gives:
\begin{equation}
H=\sum_{n \atop a=1,2} a_{n,a}^+a_{n+1,a}^- + a_{n,a}^-a_{n+1,a}^+ +
\lambda \sum_n (a_{n,2}^+a_{n+1,2}^- +
a_{n,2}^-a_{n+1,2}^+)(a_{n,1}^+a_{n+1,1}^- + a_{n,1}^-a_{n+1,1}^+)
\end{equation}

The mean field approximation is obtained by taking: (recall, $a=1,2$  is a chain index). 
\begin{equation}
\langle a_{n,a}^+a_{n+1,a}^- + a_{n,a}^-a_{n+1,a}^+ \rangle =t + (-)^{n+a}
\delta
\end{equation}

The mean field equation are then:
\begin{eqnarray}\label{eq:mean_field_symmetric}
1=\frac{2\lambda}{\pi} \sqrt{1 +\left(\frac{\delta}{1+\lambda
t}\right)^2}\left[ 2\mathbf{K}\left(1-\left(\frac{\delta}{1+\lambda
t}\right)^2\right)-\frac{\mathbf{E}\left(1-\left(\frac{\delta}{1+\lambda
t}\right)^2\right)}{1-\left(\frac{\delta}{1+\lambda
t}\right)^2}\right] \nonumber \\
t=-\frac{2}{\pi}(1+\lambda t)\sqrt{1 +\left(\frac{\delta}{1+\lambda
t}\right)^2}\frac{\mathbf{E}\left(1-\left(\frac{\delta}{1+\lambda
t}\right)^2\right)}{1-\left(\frac{\delta}{1+\lambda
t}\right)^2}
\end{eqnarray}
Where $\mathbf{K},\mathbf{E}$ are the complete elliptic integrals of
the first and second kind respectively.
For small $\delta$, the mean field equations
(\ref{eq:mean_field_symmetric}) reduce to :
\begin{eqnarray}\label{eq:symmetric_mf_small}
\frac{t}{1+\lambda t}&=&-\frac 2 \pi \nonumber \\
1&=&\frac {4\lambda}{\pi} \ln \left( \frac{\sqrt{8} \mid 1 +\lambda t
\mid}{\mid \delta \mid}\right)
\end{eqnarray}

And one obtains:
\begin{eqnarray}
t&=&-\frac{2}{\pi+2\lambda} \nonumber \\
\delta&=&\sqrt{8} \frac{\pi - 2\lambda}{\pi +2 \lambda}\exp
\left(-\frac{\pi}{4\lambda}\right)
\end{eqnarray}

This result from the mean field theory agrees with the prediction of
bosonization since bosonization would predict the same essential
singularity in $\delta$ at small $\lambda$, the interchain coupling
being marginal.

\subsection{Non equivalent XY chains}

In this section, we consider a problem with a Hamiltonian of the form:
\begin{equation}
H=\sum_n (S_n^+S_{n+1}^- + S_n^-S_{n+1}^+)(1 +\alpha  (\tau_n^+\tau_{n+1}^- +
\tau_n^-\tau_{n+1}^+))
\end{equation}
After a Jordan Wigner transformation, the Hamiltonian becomes:
\begin{equation}
H=\sum_n (a_n^+a_{n+1}^- + a_n^-a_{n+1}^+)(1 +\alpha  (b_n^+b_{n+1}^- +
b_n^-b_{n+1}^+))
\end{equation}

The mean field approximation is now:
\begin{eqnarray}
\langle a_n^+a_{n+1}^- + a_n^-a_{n+1}^+\rangle = t_1 + (-)^n \delta_1 \\
\langle b_n^+b_{n+1}^- + b_n^-b_{n+1}^+\rangle = t_2 + (-)^n \delta_2
\end{eqnarray}

The mean field equations are now:
\begin{eqnarray}
t_1=-\frac{2}{\pi}(1+\alpha t_2)\sqrt{1 +\left(\frac{\delta_2}{1+\alpha
t_2}\right)^2}\frac{\mathbf{E}\left(1-\left(\frac{\delta_2}{1+\alpha
t_2}\right)^2\right)}{1-\left(\frac{\delta_2}{1+\alpha
t_2}\right)^2} \nonumber \\
\delta_1=-\frac{2\alpha \delta_2}{\pi}\sqrt{1 +\left(\frac{\delta_2}{1+\alpha
t_2}\right)^2}\left[ 2\mathbf{K}\left(1-\left(\frac{\delta_2}{1+\alpha
t_2}\right)^2\right)-\frac{\mathbf{E}\left(1-\left(\frac{\delta_2}{1+\alpha
t_2}\right)^2\right)}{1-\left(\frac{\delta_2}{1+\alpha
t_2}\right)^2}\right] \nonumber \\
t_2=-\frac{2}{\pi}(1+\alpha t_2)\sqrt{1
+\left(\frac{\delta_1}{t_1}\right)^2}\frac{\mathbf{E}\left(1-\left(\frac{\delta_1}{t_1}\right)^2\right)}{1-\left(\frac{\delta_1}{t_1}\right)^2}
\nonumber \\
\delta_2=-\frac{2\alpha \delta_1}{\pi}\sqrt{1 +\left(\frac{\delta_1}{
t_1}\right)^2}\left[
2\mathbf{K}\left(1-\left(\frac{\delta_1}{t_1}\right)^2\right)-\frac{\mathbf{E}\left(1-\left(\frac{\delta_1}{t_1}\right)^2\right)}{1-\left(\frac{\delta_1}{t_1}\right)^2}
\right] \\
\end{eqnarray}

for small $\alpha$, the mean field equations for $t_1$ and $\delta_1$
reduce to:
\begin{eqnarray}
t_1=-\frac 2 \pi \nonumber \\
\delta_1=-\frac{4\alpha \delta_2 }{\pi}\ln \left(\frac {\sqrt{8} \mid 1+\alpha
t_2\mid }{\mid \delta_2 \mid}\right) \nonumber \\
\end{eqnarray}
One sees that a finite bandwidth $2\alpha/\pi$ is produced for the $\tau$ spin
waves
at least for small $\alpha$. The gap equations admit solutions at
small $\delta$.
One expects at weak coupling an essential singularity in $\delta_1
\sim \exp(-Cte/\alpha)$. Therefore, $\delta_1/t_1$ should go to zero
for $\alpha \to 0$. This implies that one can write:
\begin{eqnarray}
t_2=-\frac{2\alpha t_1}{\pi} \\
\delta_2=-\frac{2\alpha \delta_1}{\pi} \ln \left(\sqrt{8} \left| \frac
{t_1}{\delta_1}\right| \right)
\end{eqnarray}
One can solve the mean field equations for $\delta_1$ and $\delta_2$.
One obtains:
\begin{eqnarray}
\delta_1=\frac{4}{\sqrt{\pi}}(1+o(1))\exp\left(-\frac{\pi}{4\alpha}\right)
\nonumber \\
\delta_2=-\frac{4}{\sqrt{\pi}}(1+o(1))\exp\left(-\frac{\pi}{4\alpha}\right)  \\
\end{eqnarray}

We obtain therefore self-consistently a gap much smaller than the
smallest bandwidth. As a consequence, the correlation length is much
larger than the lattice spacing, which justifies a continuum
approximation. The bosonization treatment is therefore valid for
$\alpha \to 0$ and close to the XY limit.

\section{Renormalization group equations derived by momentum shell
integration}\label{app:momentum_shell_rg}
\subsection{The Quantum sine Gordon Model}

In this section, we derive RGE for the quantum sine Gordon model using
the method of Knops and Den Ouden \cite{knops_sine-gordon}. 
The quantum sine Gordon model is defined by the following lattice
Hamiltonian:
\begin{eqnarray}
H= \sum_i \left\{\frac{v_F a}{2\pi} \left[ (\pi \Pi_i)^2 +
(\phi_{i+1}-\phi_i)^2\right] - \frac{2g}{(2\pi a)^2}  \cos 4\phi_i \right\}
\end{eqnarray} 
The cutoff in this model is only on space and not on time.

The Euclidean action of the Quantum Sine Gordon model is obtained as: 
\begin{eqnarray}
S_E[\phi]=\int_{-\infty}^{\infty} \frac{d\omega}{2\pi} \int_{-\frac
\pi a}^{\frac \pi a} \frac{dk}{\pi} \frac{u}{2\pi K}
\left(k^2+\frac{\omega^2}{u^2} \right) |\phi(k,\omega)|^2
-\frac{2g}{(2\pi a)^2} \int dx d\tau \cos 4\phi
\end{eqnarray}

Where $\int dx \to a \sum_i$. One can integrate from $-\infty$ to
$\infty$ using a cutoff function $\varphi(k)=\Theta(\frac \pi a
-|k|)$. This action breaks rotation invariance in the $x,\tau$ space
in contrast with the action of the classical sine-Gordon model
\cite{knops_sine-gordon}. Instead of the sharp cutoff, one can use any
cutoff function $\varphi(ak)$ such that $\varphi(0)=1$ and $\lim_{x \to
\infty} \varphi(x)=\infty$. An example is $\varphi(ak)=e^{-a |k|}$.

the technique developed by Knops and Den
Ouden\cite{knops_sine-gordon} for the classical 
sine-Gordon model is straightforwardly adapted to
the Quantum Sine-Gordon  model. One obtains:

\begin{eqnarray}\label{eq:rge_momentum_cutoff}
\frac{dg}{dl}=(2-4K)g \\
\frac{d}{dl}\left(\frac u K \right) = u \left(\frac g {\pi
u}\right)^2 \int_0^\infty d\rho \rho^{4(1-2K)} \frac{\partial
F_2}{\partial \rho}(\rho) \\
\frac{d}{dl}(uK)= \frac 1 u \left(\frac g {\pi
u}\right)^2 \int_0^\infty d\rho \rho^{4(1-2K)} \frac{\partial
F_1}{\partial \rho}(\rho) 
\end{eqnarray}
Where:
\begin{eqnarray}
F_1(\rho)=\int_0^{2\pi}\sin^2 \theta e^{16 K \left[V(\rho, \theta)+ \frac 1 2 \ln \rho\right]}\frac {d \theta}{\pi}
\nonumber \\
F_2(\rho)=\int_0^{2\pi}\cos^2 \theta e^{16 K \left[V(\rho,\theta)+ \frac 1 2 \ln \rho\right]}\frac {d \theta}{\pi}
\nonumber \\
\end{eqnarray}
And:
\begin{equation}
V(\rho,\theta)=\frac 1 2 \int_0^\infty \frac {d\kappa}{\kappa}
\varphi(\kappa)(\cos (\kappa \rho \cos \theta)e^{-\kappa \rho | \sin
\theta |}-1)
\end{equation}

For instance, if $\phi(\kappa)=e^{-\kappa}$, one has:
\begin{equation}
V(\rho,\theta)=-\frac 1 2 \ln \left(\frac {\sqrt{\rho^2 \cos^2\theta
+(1+\rho |\sin \theta |)^2}} a \right)
\end{equation}

The Eqs. (\ref{eq:rge_momentum_cutoff}) have to be contrasted with the
usual RG equations for the sine Gordon model with a cutoff isotropic
in the $x,u\tau$ space. In the latter case, the equations are such
that $du/dl=0$, and only $K$ is flowing under renormalization. 
In the case we have considered, both $u$ and $K$ are flowing under RG
transformation. 
However, for $K=1/2$, the RG equations (\ref{eq:rge_momentum_cutoff})
are considerably simplified. Since $F_1(0)=F_2(0)=0$ and
$F_1(\infty)=F_2(\infty)=1$, one has the following simplified RG
equations:
\begin{eqnarray}
\frac{du}{dl}=0\nonumber \\
\frac{dK}{dl}=\left(\frac g {\pi u}\right)^2 \nonumber \\
\frac {dg}{dl}=(2-4K) g
\end{eqnarray}

These RG equations are identical to the ones obtained with a cutoff
isotropic in $x,u\tau$ space. we conclude that in the vicinity of the
BKT transition point, the anisotropy between space and time does not
matter. 

\subsection{Renormalization of the dimerization term}

We consider the case where $\delta_1=\delta_2=0$, with Euclidean action:
\begin{eqnarray}
S=\sum_{\alpha=1,2} \int dx d\tau \left[\frac{u_\alpha (\partial_x
\phi_\alpha)^2 }{2\pi K_\alpha} +\frac{(\partial_\tau
\phi_\alpha)^2}{2\pi u_\alpha K_\alpha} -\frac{2g}{(2\pi a)^2}\sin
2\phi_1 \sin 2\phi_2 \right]
\end{eqnarray}

Using the method of Knops and Den Ouden, we obtain the following RG
equations:
\begin{eqnarray}\label{eq:rge_biquad}
\frac{dg}{dl}=(2-K_1-K_2) g \nonumber \\
\frac d {dl}\left( \frac {u_1}{K_1} \right)=\frac {u_1}{8}
\left(\frac{g}{\pi u_1}\right)^2 \int_0^{\infty}d\rho
\rho^{2(2-K_1-K_2)} \partial_\rho F_3(\rho)\nonumber \\
 \frac d {dl}\left( \frac 1 {u_1K_1} \right)=\frac {1}{8u_1}
\left(\frac{g}{\pi u_1}\right)^2 \int_0^{\infty}d\rho
\rho^{2(2-K_1-K_2)} \partial_\rho F_4(\rho)\nonumber \\
\frac d {dl}\left( \frac {u_2}{K_2} \right)=\frac {u_2}{8}
\left(\frac{g}{\pi u_2}\right)^2 \int_0^{\infty}d\rho
\rho^{2(2-K_1-K_2)} \partial_\rho F_5(\rho)\nonumber \\
 \frac d {dl}\left( \frac 1 {u_2K_2} \right)=\frac {1}{8u_2}
\left(\frac{g}{\pi u_2}\right)^2 \int_0^{\infty}d\rho
\rho^{2(2-K_1-K_2)} \partial_\rho F_6(\rho)
\end{eqnarray}

Where:

\begin{eqnarray}
F_3(\rho)=\int_0^{2\pi}\frac {d\theta}{ \pi} \cos^2 \theta \exp \left( 4 K_1 V(\rho,\theta,1)+ 4 K_2 V(\rho, \theta, \frac {u_2}{u_1})
+2(K_1+K_2) \ln \rho \right) \nonumber \\
F_4(\rho)=\int_0^{2\pi}\frac {d\theta}{ \pi} \sin^2 \theta \exp \left( 4 K_1 V(\rho,\theta,1)+ 4 K_2 V(\rho, \theta, \frac {u_2}{u_1})
+2(K_1+K_2) \ln \rho\right) \nonumber \\
F_5(\rho)=\int_0^{2\pi}\frac {d\theta}{ \pi} \cos^2 \theta \exp \left( 4 K_1 V(\rho,\theta,\frac{u_1}{u_2})+ 4 K_2 V(\rho, \theta, 1)
+2(K_1+K_2) \ln \rho\right) \nonumber \\
F_4(\rho)=\int_0^{2\pi}\frac {d\theta}{ \pi} \sin^2 \theta \exp \left( 4 K_1 V(\rho,\theta,\frac{u_1}{u_2})+ 4 K_2 V(\rho, \theta, 1)
+2(K_1+K_2) \ln \rho\right) 
\end{eqnarray}

And:
\begin{eqnarray}
V(\rho,\theta,\alpha)=\frac 1 2 \int_0^\infty \frac {d\kappa}{\kappa}
\varphi(\kappa)(\cos (\kappa \rho \cos \theta)e^{-\kappa \alpha \rho | \sin
\theta |}-1)
\end{eqnarray}

It is easily seen that similarly to the case in which $K_1+K_2=2$,
universal coefficients appear in the RG equations. These universal
coefficients are respectively:
\begin{eqnarray}
F_3(\infty)=\int_0^{2\pi} \frac{\cos^2 \theta}{\left(\cos^2\theta
+\left(\frac{u_2}{u_1}\right)^2 \sin^2 \theta\right)^{K_2}}
\frac{d\theta}{\pi} \nonumber \\
F_4(\infty)=\int_0^{2\pi} \frac{\sin^2 \theta}{\left(\cos^2 \theta
+\left(\frac{u_2}{u_1}\right)^2 \sin^2 \theta\right)^{K_2}}
\frac{d\theta}{\pi} \nonumber  \\
F_5(\infty)=\int_0^{2\pi} \frac{\cos^2 \theta}{\left(\cos^2 \theta
+\left(\frac{u_1}{u_2}\right)^2 \sin^2 \theta\right)^{K_1}}
\frac{d\theta}{\pi} \nonumber \\
F_6(\infty)= \int_0^{2\pi} \frac{\sin^2 \theta}{\left(\cos^2 \theta
+\left(\frac{u_1}{u_2}\right)^2 \sin^2  \theta\right)^{K_1}}
\frac{d\theta}{\pi} \\
\end{eqnarray}

One can check easily that if $u_1=u_2$, the RG equations
(\ref{eq:rge_biquad}) reduce to
\begin{eqnarray}
\frac{d}{dl}\left(\frac 1 {K_1}\right)=\frac{g^2}{8\pi^2 u_1^2}\nonumber \\
\frac{d}{dl}\left(\frac 1 {K_2}\right)=\frac{g^2}{8\pi^2 u_2^2} \\
\end{eqnarray}
which is equation (3.24) in the spintube paper for $\delta_1=\delta_2=0$.

\subsection{The full problem}

For the full problem, the Lagrangian is:
\begin{eqnarray}
L& = & \int dx d\tau \left[\frac{u_1 (\partial_x\phi_1)^2}{2\pi K_1}
+\frac{(\partial_\tau\phi_1)^2}{2\pi u_1 K_1} -\frac{2\delta_1}{(2\pi
a)^2} \cos 4 \phi_1  \right.\nonumber \\
& + &  \frac{u_2 (\partial_x\phi_2)^2}{2\pi K_2}
+\frac{(\partial_\tau\phi_1)^2}{2\pi u_2 K_2} -\frac{2\delta_2}{(2\pi
a)^2} \cos 4 \phi_2  \nonumber \\
& + &\left. \frac{2g}{(2\pi a)^2} 2 \sin 2\phi_1 \sin 2\phi_2 \right]
\end{eqnarray}
Applying the Knops Den Ouden Method, we obtain the following
Renormalization Group equations: 
\begin{eqnarray}
\frac{d}{dl}\left(\frac{u_1}{K_1}\right)& =&\frac{u_1}{8}\left(\frac{g}{\pi
u_1}\right)^2 \int_0^\infty d\rho \rho^{2(2-K_1-K_2)}\partial_\rho F_3
+u_1 \left(\frac{\delta_1}{\pi u_1}\right)^2\int_0^\infty d\rho
\rho^{4(1-2K_1)} \partial_\rho F_2 \nonumber \\
\frac{d}{dl}\left(\frac{1}{u_1K_1}\right)&=&\frac{1}{8u_1}\left(\frac{g}{\pi
u_1}\right)^2 \int_0^\infty d\rho \rho^{2(2-K_1-K_2)}\partial_\rho F_4
+\frac 1 {u_1} \left(\frac{\delta_1}{\pi u_1}\right)^2\int_0^\infty d\rho
\rho^{4(1-2K_1)} \partial_\rho F_1 \nonumber \\
\frac{d}{dl}\left(\frac{u_2}{K_2}\right)&=&\frac{u_2}{8}\left(\frac{g}{\pi
u_2}\right)^2 \int_0^\infty d\rho \rho^{2(2-K_1-K_2)}\partial_\rho F_5
+u_2 \left(\frac{\delta_2}{\pi u_2}\right)^2\int_0^\infty d\rho
\rho^{4(1-2K_2)} \partial_\rho F_2 \nonumber \\
\frac{d}{dl}\left(\frac{1}{u_2 K_2}\right)&=&\frac{1}{8u_2}\left(\frac{g}{\pi
u_2}\right)^2 \int_0^\infty d\rho \rho^{2(2-K_1-K_2)}\partial_\rho F_6
+\frac 1 {u_2} \left(\frac{\delta_2}{\pi u_2}\right)^2\int_0^\infty d\rho
\rho^{4(1-2K_2)} \partial_\rho F_1 \nonumber \\
\frac{d}{dl}\left(\frac {\delta_1}{\pi u_1}\right)&=&(2-4K_1)\frac
{\delta_1}{\pi u_1} -\frac 1 8 \left(\frac{g}{\pi
u_1}\right)^2\int_0^\infty d\rho \rho^{2(1+K_1-K_2)}\partial_\rho F_8
\nonumber \\
\frac{d}{dl}\left(\frac {\delta_2}{\pi u_2}\right)&=&(2-4K_2)\frac
{\delta_2}{\pi u_2} -\frac 1 8 \left(\frac{g}{\pi
u_2}\right)^2\int_0^\infty d\rho \rho^{2(1+K_2-K_1)}\partial_\rho F_7
\\
\frac{d}{dl}\left(\frac {g}{\pi u_1}\right)&=&(2-K_1- K_2)\frac
{g}{\pi u_1}-\frac 1 2 \frac g {\pi u_1}\left(\frac {\delta_1}{\pi
u_1}\int_0^\infty d\rho \rho^{2-4K_1}\partial_\rho F_9 +\frac {\delta_2}{\pi
u_2}\int_0^\infty d\rho \rho^{2-4K_2}\partial_\rho F_9\right) \nonumber
\end{eqnarray}

Where:
\begin{eqnarray}
F_7(\rho)=\int_0^{2\pi} \frac{d\theta}{2\pi} e^{4\left[K_1
V(\rho,\theta,\frac{u_1}{u_2}) - K_2
V(\rho,\theta)\right]-2(K_1-K_2)\ln \rho} \nonumber \\
F_8(\rho)=\int_0^{2\pi} \frac{d\theta}{2\pi} e^{4\left[K_1
V(\rho,\theta) - K_2
V(\rho,\theta,\frac{u_2}{u_1})\right]-2(K_2-K_1)\ln \rho} \nonumber \\
F_9(\rho)=\int_0^\infty e^{8K_1 \left[V(\rho,\theta)+\frac 1 2 \ln \rho\right]}
\end{eqnarray}

One can check that if $u_1=u_2=u$, these equations reduce to those
derived using OPE techniques. 

Clearly, the presence of a finite velocity difference results in
different coefficients in the RG equations. However, this should not
affect the topology of the phase diagram.

\newpage

\narrowtext

\begin{figure}
\centerline{\epsfig{file=fig4.ps,angle=0,width=12cm}}
\caption{ (a) magnetic soliton. These solitons are associated with the
triplet excitations having $m=\pm 1$. On the figure, $m=-1$;
(b) non-magnetic soliton. These solitons are associated with the
triplet excitation having $m=0$ or the singlet excitation.}
\label{fig:solitons}
\end{figure}

\begin{figure}
\centerline{\epsfig{file=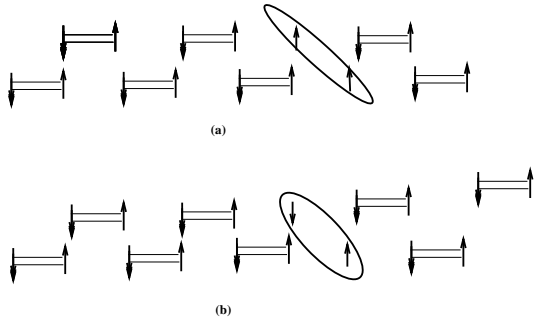,angle=0,width=7cm}}
\vspace{0.5 cm}
\caption{(a) triplet excitation of the spin ladder with
biquadratic exchange;
(b) singlet excitation of the spin ladder with
biquadratic exchange}
\label{fig:singlet_triplet}
\end{figure}

\end{document}